\documentclass[10pt,journal]{IEEEtran}
\IEEEoverridecommandlockouts
\makeatletter
\def\endthebibliography{%
	\def\@noitemerr{\@latex@warning{Empty `thebibliography' environment}}%
	\endlist
}
\makeatother
\usepackage{cite}
\usepackage{amsthm,amssymb,amsfonts}
\usepackage{amsmath}
\usepackage{graphicx}
\usepackage{xcolor}
\usepackage{array}
\usepackage{verbatim}
\usepackage{multirow}
\usepackage{booktabs}
\usepackage{makecell}
\usepackage{balance}
\usepackage{float}
\usepackage{subfigure}
\usepackage{enumerate}

\usepackage[T1]{fontenc}
\usepackage[edges]{forest}

\def\BibTeX{{\rm B\kern-.05em{\sc i\kern-.025em b}\kern-.08em 
		T\kern-.1667em\lower.7ex\hbox{E}\kern-.125emX}}
\usetikzlibrary{arrows,shapes,positioning,shadows,trees}
\usepackage{tikz}
\tikzstyle{every node}=[draw=black,anchor=west,rounded corners=0.05cm,font=\footnotesize,drop shadow]

\PassOptionsToPackage{hyphens}{url}
\usepackage{hyperref}
\hypersetup{colorlinks, citecolor=green, filecolor=pink, linkcolor=red, urlcolor=blue }

\begin{document}
	\title{Unmanned Aerial Vehicle for Internet of Everything: Opportunities and Challenges}
	\author{Yalin Liu, Hong-Ning Dai, ~\IEEEmembership{Senior Member, IEEE}, Qubeijian Wang, Mahendra K. Shukla, Muhammad Imran
		\thanks{Y. Liu, H.-N. Dai and Q. Wang are with the Faculty of Information Technology, Macau University of Science and Technology, Macau SAR (email: yalin\_liu@foxmail.com; hndai@ieee.org; qubeijian.wang@gamil.com).}
		\thanks{M. K. Shukla is with the Department of Electrical and Computer Engineering, University of Saskatchewan, Canada (email: m.shukla@usask.ca).}
		\thanks{M. Imran is with the College of Applied Computer Science, King Saud University, Riyadh, Saudi Arabia. (email: dr.m.imran@ieee.org).}
	}

	\maketitle
	\begin{abstract}
		The recent advances in information and communication technology (ICT) have further extended Internet of Things (IoT) from the sole ``things'' aspect to the omnipotent role of ``intelligent connection of things''. Meanwhile, the concept of internet of everything (IoE) is presented as such an omnipotent extension of IoT. However, the IoE realization meets critical challenges including the restricted network coverage and the limited resource of existing network technologies. Recently, Unmanned Aerial Vehicles (UAVs) have attracted significant attentions attributed to their high mobility, low cost, and flexible deployment. Thus, UAVs may potentially overcome the challenges of IoE. This article presents a comprehensive survey on opportunities and challenges of UAV-enabled IoE. We first present three critical expectations of IoE: 1) scalability requiring a scalable network architecture with ubiquitous coverage, 2) intelligence requiring a global computing plane enabling intelligent things, 3) diversity requiring provisions of diverse applications. Thereafter, we review the enabling technologies to achieve these expectations and discuss four intrinsic constraints of IoE (i.e., coverage constraint, battery constraint, computing constraint, and security issues). We then present an overview of UAVs. We next discuss the opportunities brought by UAV to IoE. Additionally, we introduce a UAV-enabled IoE (Ue-IoE) solution by exploiting UAVs's mobility, in which we show that Ue-IoE can greatly enhance the scalability, intelligence and diversity of IoE. Finally, we outline the future  directions in Ue-IoE.
		
	\end{abstract}
	
	\section{Introduction}
	\label{sec:intro}
	\IEEEPARstart{I}{nternet} of Everything (IoE) represents a fantastic vision in the future, in which \textit{everything} is connected to the Internet, thereby offering intelligent services and facilitating decision-making \cite{miraz2015review}. IoE's implementation depends on interdisciplinary technical innovations such as sensor and embedded technologies, low power communications and big data analytics \cite{holland2019internet}. In decades of years, the increasing technical innovations are emerged to offer new bricks to build IoE. First, the advances in sensor and embedded technologies have made the Internet of Things (IoT) nodes being more portable and less energy consumption \cite{RN109,UAVdata2018,holland2019internet}. Second, the appearance of Low Power Wide Area Network (LPWAN) technologies enables the ubiquitous network connections of low power IoT nodes \cite{holland2019internet}. Furthermore, the breakthrough in artificial intelligence and the availability of massive IoT data have driven the intelligence of IoE. In this way, IoE can be applied in wide applications such as smart manufacturing, smart agriculture and intelligent transportation system. 
	
	In 2012, CISCO has used the term ``IoE'' to envision the promising future of the Internet \cite{cisco2012}. In CISCO's view, IoE is built upon the ``four pillars'' in terms of \emph{people}, \emph{data}, \emph{process}, and \emph{things}. In contrast, IoT only contains the pillar of ``things'' \cite{miraz2015review,hussain2017internet}. Clearly, IoE extends the \emph{connection-of-machine} capability of IoT, thereby to aid \emph{automated and people-based processes} for all the \emph{things}. Compared with IoT, IoE is more insightful, i.e., enriching the lives of people by enabling all business-and-industrial processes automated and smart. To achieve this goal, IoE is desired to satisfy three expectations: 1) \textit{scalability} means to establish a scalable network architecture with ubiquitous coverage; 2) \textit{intelligence} implies to enable intelligent decisions and actions for all devices in IoE; 3) \textit{diversity} indicates supporting diverse applications. Therefore, the realization of IoE essentially depends on the achievement of the above three expectations.
	
	In recent of years, we have witnessed the rapid development of ICT technologies that can facilitate the realization of IoE. In particular, ICT technologies have further extended existing human-oriented Internet to machine-oriented Internet of \textit{Things} \cite{mekki2019comparative}, which consists of wireless sensor networks (WSN) for connecting multiple sensor nodes via an self-organized topology \cite{khan2015wireless}, low power wide area network (LPWAN) for offering large-range coverage of power-constrained nodes \cite{RN68,mekki2019comparative}, and 4G and 5G mobile networks for supporting massive-access services of machine-to-machine (M2M) communications \cite{RN12}. Meanwhile, massive data are generated from various things in real-time manner. The breakthrough of artificial intelligence (AI) technologies integrating with massive IoT data brings the opportunities in realizing intelligent applications including intelligent recognition, intelligent management and intelligent decision \cite{dai2019big}. As a result, conventional IoT has been evolved into IoE that supports intelligent connection of things, thereby enabling smart applications, such as smart meter \cite{kim2017smart}, smart grid, smart manufacturing \cite{dai2019manu}, smart agriculture, intelligent traffic scheduling \cite{RN11}. In this regard, the existing wireless communication networks (including WSN, LPWAN and 5G cellular networks) can help to realize the scalability of IoE. Meanwhile, the cutting-edge big data analysis technologies and artificial intelligence can be used for enabling intelligence to IoE. Finally, the emerging IoE intelligent services have proliferated a huge market for diverse IoE applications.
	
	However, there are still a number of intrinsic limitations preventing IoE from achieving the above three expectations. In particular, IoE has network coverage/access constraints, battery constraint of IoE nodes, security and privacy vulnerabilities. First, the existing network infrastructures are \emph{coverage-constrained} in some harsh and remote geographical areas due to the restricted deployment of network infrastructures. Hence the ubiquitous connections of IoE cannot be achieved. In addition, IoE nodes also suffer from the limited battery capacity due to cost and portability considerations. As a result, the battery-constrained nodes are easily exhausted and eventually lead to the connection-lost. This case is especially severe for the nodes in the coverage-constrained areas. Furthermore, most of the nodes in IoE have no enough computing capability to process local sensor data. Moreover, the over-simplified access protocols (e.g., NB-IoT and LPWAN) also pose potential security vulnerabilities in IoE, such as information being eavesdropped or being forged by malicious relay nodes. To address these challenges, IoE requires a flexible-coverage and elastic-deployment so as to achieve the ubiquitous coverage and offer quick response in a highly-efficient and reliable way.
	
	\begin{figure*}[t]
		\centerline{\includegraphics[width=15cm]{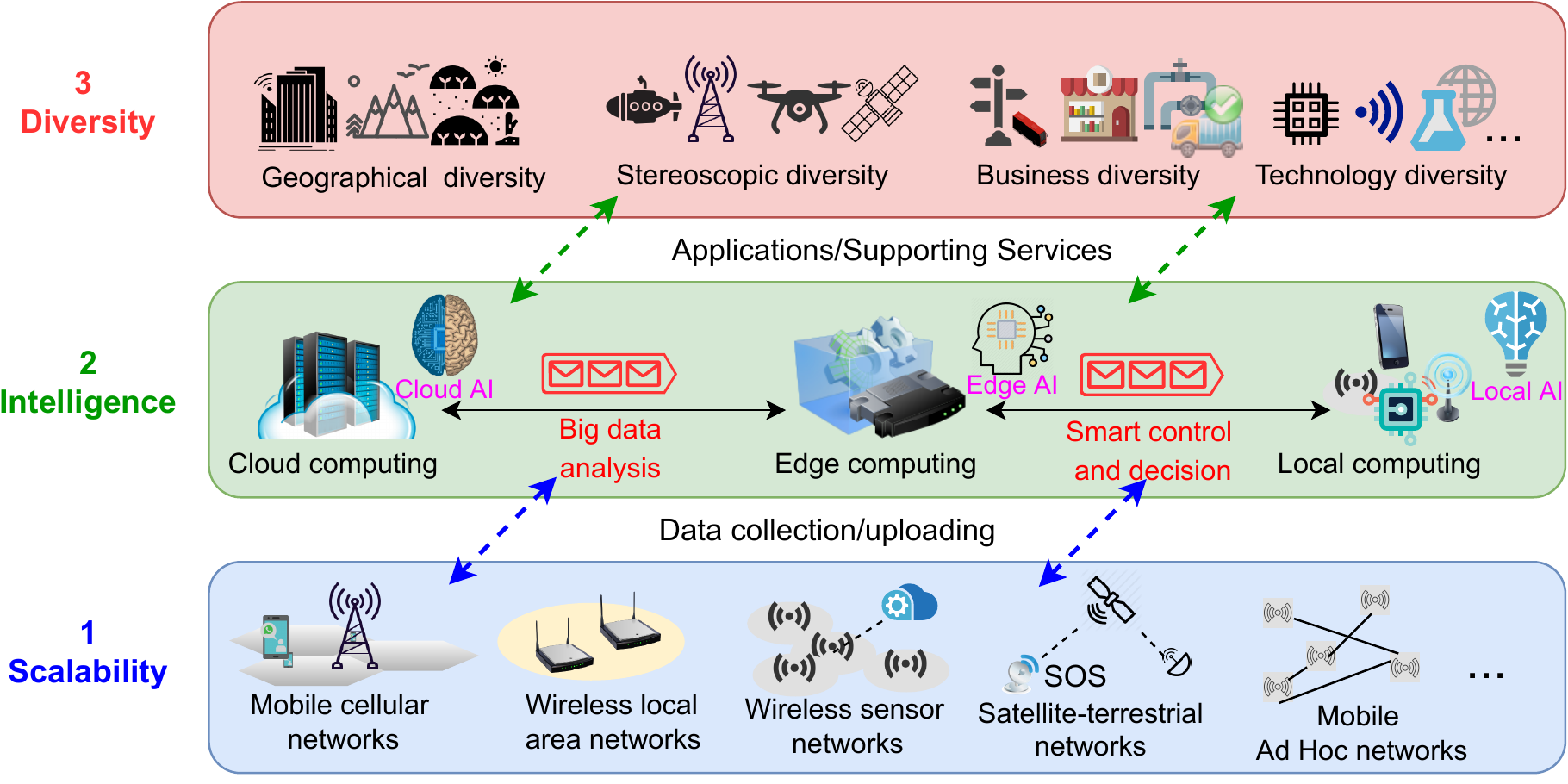}}
		\caption{Three expectations of IoE (i.e., scalability, intelligence, and diversity).}
		\label{fig: expectations}
	\end{figure*}
	
	In recent decades, Unmanned Aerial Vehicles (UAVs) have attracted significant attentions from both industrial and academia communities due to their flexible and elastic services. With high mobility and elastic deployment, UAVs can extend the coverage of IoE \cite{RN61,RN59}. In particular, compared with terrestrial networks and satellite remote communications, low-altitude UAVs-enabled wireless networks can be quickly deployed and be flexibly reconfigured to enhance the network coverage and capacity~\cite{RN59}. Furthermore, the flexible deployment of UAVs also enables myriad IoE applications. Through the dedicated configuration and specified control design, UAVs have a variety of applications such as agriculture management~\cite{wu2019agricultural}, marine mammals monitoring \cite{aniceto2018monitoring}, and military medical evacuation \cite{Handford2018Prospective}. However, using UAV in IoE also poses new challenges in the perspectives of dynamic network connection, flexible network topology, precise control and lightweight intelligent algorithms. In order to address these challenges, we need to design the dedicated communication protocols, the flexible resource allocation mechanism, the optimal trajectory plans of UAVs and the portable intelligent algorithms in IoE.
	
	Although several surveys have already discussed the IoE vision in recent decades \cite{jara2013internet,hussain2016cmos,abdelwahab2014enabling,8707052}, they mainly focused on a single IoE technology. For example, the authors in \cite{jara2013internet} have considered the Internet Protocol version 6 (IPv6) to realize the ubiquitous communication access to the Internet while they ignored the issues about wide coverage and massive accesses of wireless networks. Furthermore, the authors in  \cite{abdelwahab2014enabling} have focused on cloud plane to realize the virtualized data services for IoE whereas they ignored the orchestration with physical networks and diverse IoE intelligent algorithms. To the best of our knowledge, the previous literature lacks the comprehensive investigation of IoE in three key expectations: scalability, intelligence and diversity. 
	
	The limitations of existing surveys motivate us to conduct a more comprehensive survey exploring IoE's expectations and enabling technologies, along with the extensive applications of UAV-enabled IoE. In particular, this survey has three major contributions: i) summarizing the key expectations of IoE (i.e., scalability, intelligence, and diversity) and reviewing the corresponding enabling technologies; ii) giving an introduction of UAVs technologies (such as maneuverability, communication and other relevant technologies) and discussing the opportunities that UAVs can bring to IoE; iii) integrating UAVs with the existing IoE's enabling technologies so as to present a UAV-enabled IoE solution. 
	
	The rest of this paper is organized as follows. Section \ref{sec:ioe} introduces IoE regarding its three expectations, enabling technologies and challenges. Section \ref{sec:uav} reviews UAV's related work: unmanned aerial system (UAS) and UAV communication networks. In Section \ref{sec:uavioe}, we introduce a UAV-enabled IoE (Ue-IoE) solution by combining UAV's and the existing ICT technologies, in which we show that Ue-IoE can greatly enhance scalability, intelligence, and diversity of IoE. Further, we outline crucial issues in Ue-IoE as well as future  directions in Section \ref{sec:issues}. Finally, Section \ref{sec:conclusion} concludes the paper.
	
	\section{Internet of Everything}
	\label{sec:ioe}
	
	This section first briefs the birth of IoE in Section~\ref{subsec:birth} and presents three key expectations of IoE in Section~\ref{subsec:expectations}. We then outline the enabling technologies to fulfill the expectations in Section~\ref{subsec:enabling_tech}, and next discuss the challenges in Section~\ref{subsec:challenges}.
	
	\subsection{The Birth of IoE}
	\label{subsec:birth}
	The emergence of IoT is undoubtedly a significant stimulation to generate the IoE concept. The term IoT, means connecting electrical or electronic devices with varying sizes and capabilities to the Internet. In recent years, primarily focusing on the connection of machine-to-machine communications, IoT technology has rapidly developed in a broad spectrum of communication protocols, networks and applications (such as 802.11ah, industrial IoT, NB-IoT) \cite{lin2017survey}. Such prosperous IoT ecosystems pave a solid foundation for the IoE's communications with a broad coverage and ubiquitous connection.
	
	However, the real birth of the IoE concept comes from the idea of enabling automated machines through ubiquitous Internet, big data processing and artificial intelligence. Back to 2012, CISCO has presented a view that IoE is built upon the ``\emph{four pillars}'' in terms of \emph{people}, \emph{data}, \emph{process}, and \emph{things} \cite{cisco2012}. This view indicates that IoE considers a comprehensive interconnection of not only ``things'' but also ``automated and people-based process'' (i.e., intelligent machines/devices). This concept goes far beyond the IoT's category of simply interconnection of ``things'' (i.e., pure machines/devices). 
	
	In addition, the proliferation of big data and AI technologies brings new \emph{bricks} for IoE's construction. In recent years, more relevant literature has replenished IoE's essence, i.e., gathering big data hiding from the Internet, in virtue of various AI algorithms, and enabling all devices/machines with the automated abilities \cite{miraz2015review,hussain2017internet,miraz2018internet,holland2019internet}. Thus, IoE has the potential to extract and analyse real-time data from millions of connected devices and then to make intelligent proactive decisions, thereby enabling ``automated intelligence''.
	
	The concept of ``IoE'' has been proposed and discussed for many years, while its realization is still remaining in its infancy. In spite of challenges in the full realization of IoE, the attractive vision of IoE will never prevent us from implementing IoE. We next discuss expectations of IoE, enabling technologies as well as challenges.

	\subsection{Three expectations of IoE}
	\label{subsec:expectations}
	The vision of IoE is to connect ubiquitous electronic devices (i.e., terminal nodes of IoE) to the Internet, then to analyze massive data generated from connected terminal nodes, and thereby to offer intelligent applications for the advancement of human society. To achieve this vision, IoE is expected to fulfill three key expectations: 1) \emph{scalability} means to establish a scalable network architecture with ubiquitous coverage; 2) \emph{intelligence} implies to construct a global computing facility enabling intelligent decisions; 3) \emph{diversity} indicates to support diverse applications. Fig.~\ref{fig: expectations} shows three expectations as well as their typical enabling technologies. In detail, we describe the three expectations as follows.
	
	\begin{enumerate}[1)]
		\item \textit{Scalability}: Scalability means to establish a scalable network for IoE to elastically cover \textit{everywhere} and \textit{everything}. In this sense,  IoE can satisfy various communication requirements for different geographical scenarios including urban, rural, underwater, terrestrial, aerial, and space. To achieve this goal, the scalable IoE network requires wide coverage, massive access, and ubiquitous connection. Such IoE networks can be built by integrating multiple communication technologies with various transmission-distance (from a few meters to a thousand meters) and different network topologies (including point-to-point topology, star topology, and hybrid topology). The underlining communication networks consist of mobile cellular networks (MCN), wireless local area networks (WLAN), WSN, satellite networks, and Mobile Ad Hoc networks (MAHN). The scalability of IoE supports the physical data collection and further provides data source for intelligent analytics.
		
		\item \textit{Intelligence}: Intelligence implies enabling intelligent analysis, predictions, decisions and actions for all devices in IoE on top of distributed computing facilities across the entire IoE. Specifically, IoE needs to collect massive data from its broad and scalable network, extract the valuable information (such as smart commands or decisions) from the collected data, and then use these information to enable intelligent actions or controls for \textit{everything}. The computing facilities consist of distributed database and  storage systems, on top of which various big data processing algorithms are  deployed. The distributed database and storage systems save the collected IoT data. Diverse big data processing algorithms include descriptive, diagnostic, predictive and prescriptive analytical schemes~\cite{dai2019manu}, which are necessary to serve different intelligent applications. With the distributed computing facilities, IoE's intelligence can be categorized into local intelligence, edge intelligence, and cloud intelligence, implying that computing facilities with the corresponding intelligent algorithms are deployed at local side (i.e., at terminal nodes), edge side and remote clouds, respectively. It is necessary to orchestrate local intelligence, edge intelligence, and cloud intelligence so as to realize IoE's global intelligence. 
		\item \textit{Diversity}: Diversity indicates diverse applications that serve the ``automated and people-based process'' of IoE. The realization of diverse IoE applications is essentially based on the scalability and intelligence of IoE since they are the prerequisites for computing capability, security, energy efficiency and network performance. Depending on a broad array of applying requirements, the ``automated and people-based process'' of IoE presents diverse classifications of applications. For instance, IoE's diversity can be categorized into: i) \emph{geographical diversity} - classified by different geographical regions, ii) \emph{stereoscopic diversity} - classified by different spatial positions, iii) \emph{business diversity} - classified by different social utilities and iv) \emph{technology diversity} - classified by different ICT technologies. In the future, with the prosperity of IoE, there will be more sorts of diversities for IoE such as intelligence diversity, equipment diversity, mobility diversity. Eventually, all these diverse applications will merge together for achieving the omnipotent IoE role.
	\end{enumerate}
	
	To achieve the above three expectations, it is necessary to deploy a large number of terminal nodes (for local sensing and control), network access nodes (providing ubiquitous connections), and computing facilities (supporting intelligence). Furthermore, diverse applications will be continuously updated with the increasing demands of various intelligent services. As a result, it is anticipated in the future that IoE needs to consume plenty of resources (including battery power, computing, storage space) to support diverse society services. Meanwhile, a serious imbalance between constrained resources and three expectations will be encountered. Hence, during the IoE realization process, the effective solution is to maximize resource utilization efficiency subjected to limited resource supply, thereby satisfying all these expectations in an on-demand manner. The on-demand manner is the fundamental design principle for the IoE enabling technologies, which will be introduced next. 
	
	\subsection{Enabling technologies for three expectations of IoE}
	\label{subsec:enabling_tech}
	\subsubsection{Enabling scalability} 
\begin{figure}[t]
		\centering{\includegraphics[width=\columnwidth]{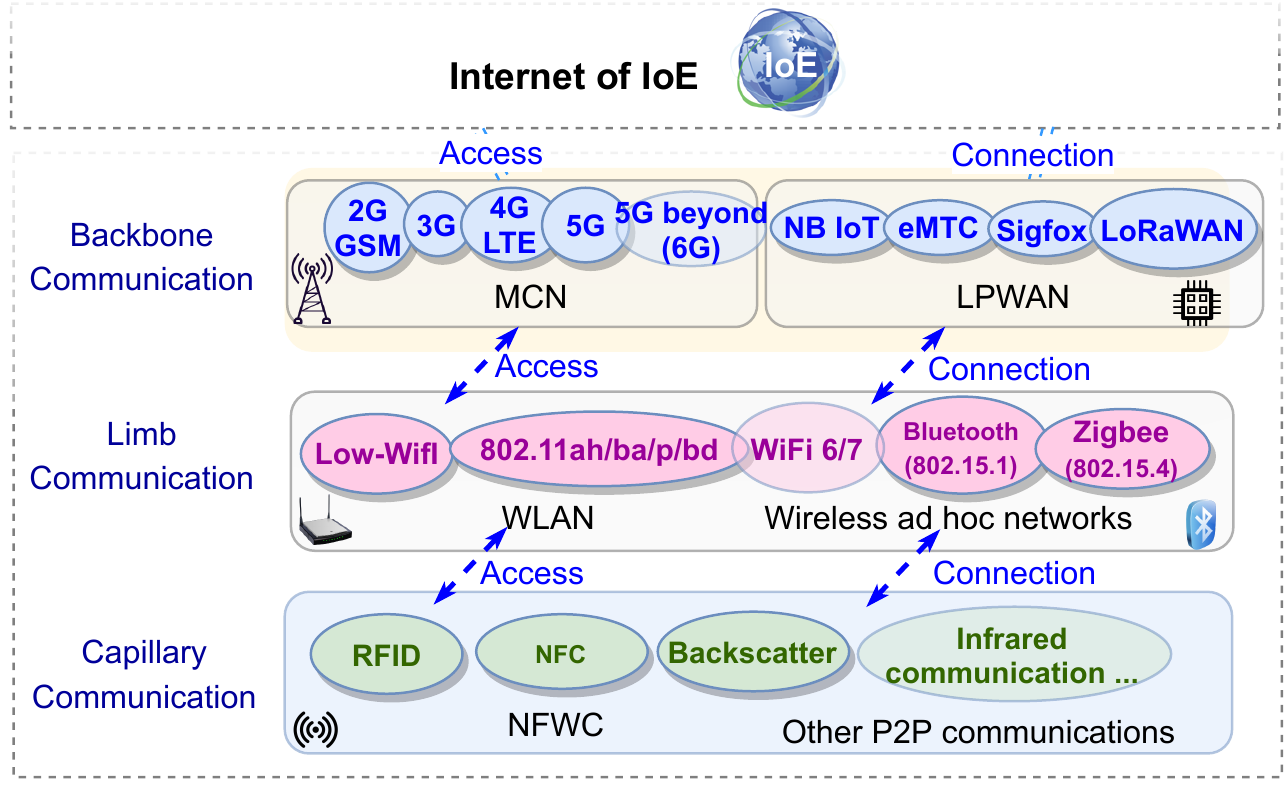}}
		\caption{Enabling communication technologies of IoE's Scalability}
		\label{fig: IOEscaena}
	\end{figure}

	Scalability for IoE means to build a global network that enables wide coverage, ubiquitous connection, and massive access. Since the network with global and ubiquitous coverage is not present, the IoE's scalability can only rely on the cooperation of various existing networks that can support different kinds of distance communications and fit for a variety of network topologies. These networks that are complementary can coordinate with each other to construct the scalable IoE. In accordance with the previous literature \cite{RN17,RN12}, we present an overview of enabling technologies to achieve IoE's scalability. The enabling technologies can be essentially categorized into three types: 1) the technologies enabling the \emph{backbone communication} of IoE (i.e., global networks), 2) the technologies enabling the \emph{limb communication of IoE} (i.e., local networks), and 3) the technologies enabling the \emph{capillary communication} of IoE (i.e., point-to-point connections), as shown in Fig. \ref{fig: IOEscaena}.

	The backbone communication of IoE is essentially to offer an overall connection of conducting data collection, transmission, processing and interaction. Hence, to offer a global coverage, the backbone communication of IoE requires the long-distance wireless communication technologies, which can support the communication distance over ten kilo-meters. The suitable candidates include incumbent mobile communication networks (MCN) and low power wide area networks (LPWANs) \cite{mekki2019comparative}. Particularly, MCN covers most densely crowd areas including business regions, and urban residential regions \cite{cosovic20175g,he2016big}. Therefore, IoE nodes can access the backbone network via the diverse MCN communication technolgoies from 2G (GPRS), 3G, 4G (LTE), 5G, and even 5G-Beyond \cite{andreev2019future}. On the other hand, LPWAN has also attracted significant attentions recent years since it cannot only provide a wide coverage but also a low-power solution for IoE \cite{RN14}. In this sense, MCNs fit for the IoE nodes with sufficient energy supply (e.g., the equipment in smart grids), while LPWANs are suitable for massive IoE nodes with power-constraints (e.g., the nodes in forest monitoring, smart agriculture). Therefore, the two technologies may complement with each other to establish the backbone communication of IoE. 
	
	Being complementary to the backbone communication, the limb communication of IoE consists of flexible local networks with a communication ranging from a few meters to hundreds of meters. The typical enabling technologies include wireless local area networks (WLAN), low power Bluetooth (BLE), Zigbee, Z-wave, 6LoWPAN and 802.11ah \cite{samuel2016review,salman2015networking}. In most cases, these limb communication technologies are used to construct Ad Hoc/mesh networks such as wireless sensor networks (WSNs) \cite{khan2015wireless}, wireless body area networks (WBANs) \cite{pramanik2019wban}, wireless personal area networks (WPANs) \cite{lee2019multichannel}. The aim of Ad Hoc/mesh networks is that all nodes can be connected together to achieve smart controls. Moreover, these networks are suitable for multi-hop communications since they can be easily used in home automation scenarios, industrial process control, body activity monitoring, indoor localization.    
	
	The capillary communication offers massive low-cost and point-to-point connections in IoE. The enabling technologies of capillary communications are mainly based on a series of near-field wireless communication (NFWC) technologies based on the inductive-coupling principle. These NFWC technologies denotes a set of communication protocols: back-scattering communication \cite{RN65}, radio-frequency identification (RFID) \cite{fernandez2017reverse}, and near-field communication (NFC) \cite{fisher2016mobile}. The data transferred in NFWC can take place from small tags to readers within a range of a few centimeters, where the tags are attached at circulated products while the reader is generally deployed at a fixed position to transmit the received data to the back-end server (to store the collected data). Hence, it can perform flexible and low-cost peer-to-peer communications. In practice, it has been widely used in myriad applications such as mobile identification systems and logistic monitoring systems \cite{fernandez2017reverse}.
	
	\begin{figure}[t]
		\centering{\includegraphics[width=\columnwidth]{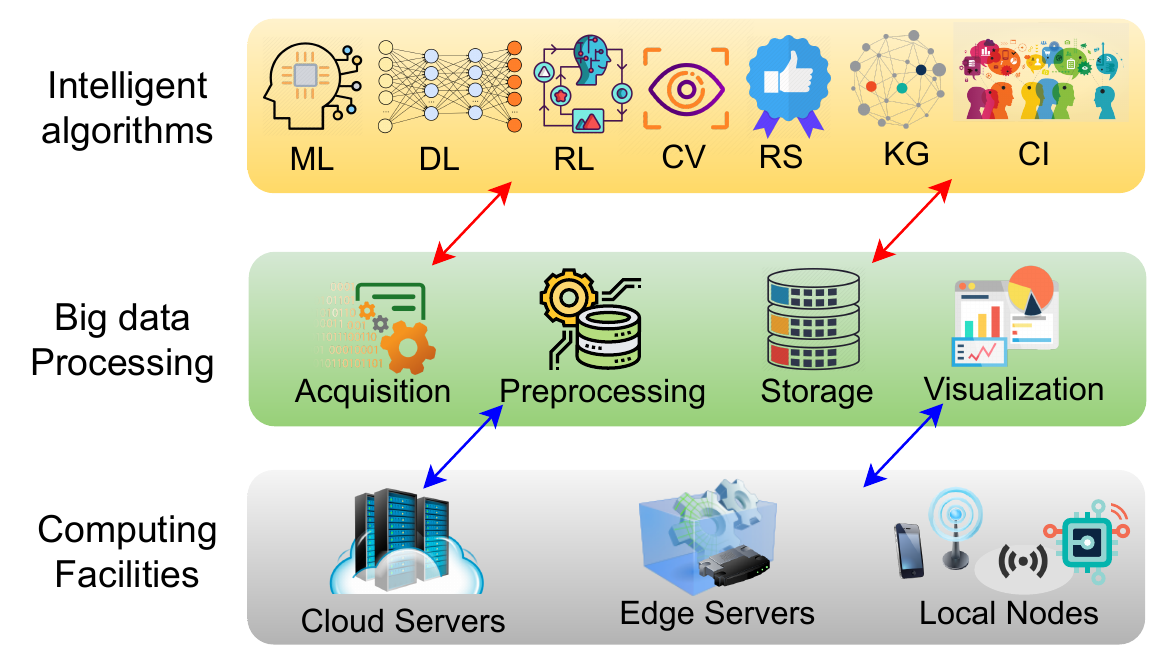}}
		\caption{Enabling technologies of IoE's Intelligence}
		\label{fig: IOEintena}
	\end{figure}	
	
	In the above enabling technologies, since the traditional networks of WCN and WLAN have been built to offer human-orientated information services, they are globally allocated with sufficient network resources for reliable information transmissions. In contrast, other enabling technologies are developed with the goals of low communication power and low hardware cost, to address the imbalance of limited network resources and massive communications between terminal nodes. This phenomenon symbolizes the on-demand principle of IoE. For example, the enabling technology of IoE's limb communication - LPWAN keeps the low power and wide coverage design principle and refers to diverse communication protocols such as LoRaWAN, ZigBee, and NB-IoT \cite{RN14}. Recently, a series of WLAN protocols have been released to support the specific IoE applications, such as 802.11ah, and 802.11p \cite{ba20161}. Additionally, low hardware cost is also a major concern for communication equipment suppliers. An example is low-cost communication chips and modules with simplified protocols stack and limited storage/battery capacity such as NB-IoT and eMTC \cite{mekki2019comparative}. 
	
	
	\subsubsection{Enabling intelligence}
	The intelligence of IoE can be enabled by performing big data processing algorithms and diverse intelligent algorithms that run on distributed computing facilities, as shown in Fig. \ref{fig: IOEintena}. Distributed computing facilities include cloud servers, edge servers, and local IoT nodes, all of which are interconnected through the backbone, limb and capillary communications. Cloud servers with abundant storage and computing resources can undertake computing-intensive or storage-intensive big data processing tasks as well as intelligent algorithms (e.g., deep learning algorithms) in a centralized manner, thereby enabling the global intelligence of IoE \cite{hwang2017big}. Edge servers possessing fewer computing resources than cloud servers are deployed at base stations, IoT gateways or access points, in approximation to users. Some less computing-intensive tasks such as data preprocessing, compression and encryption can be conducted at edge servers \cite{ren2017serving,tang2019reconfigurable}. Local nodes generally referring to the IoE nodes only have limited storage and computing resources, in which data collections or lightweight data preprocessing tasks can be conducted. Although local nodes and edge servers have less computing capabilities than cloud servers, they can process some context-aware and privacy-sensitive tasks locally. It is a necessity to orchestrate various computing resources and schedule different computing tasks at local side, edge side, and cloud side in order to enable a ubiquitous computing capability across the entire IoE, thereby realizing the global intelligence of IoE.
	
	\begin{figure}[t]
		\centering{\includegraphics[width=\columnwidth]{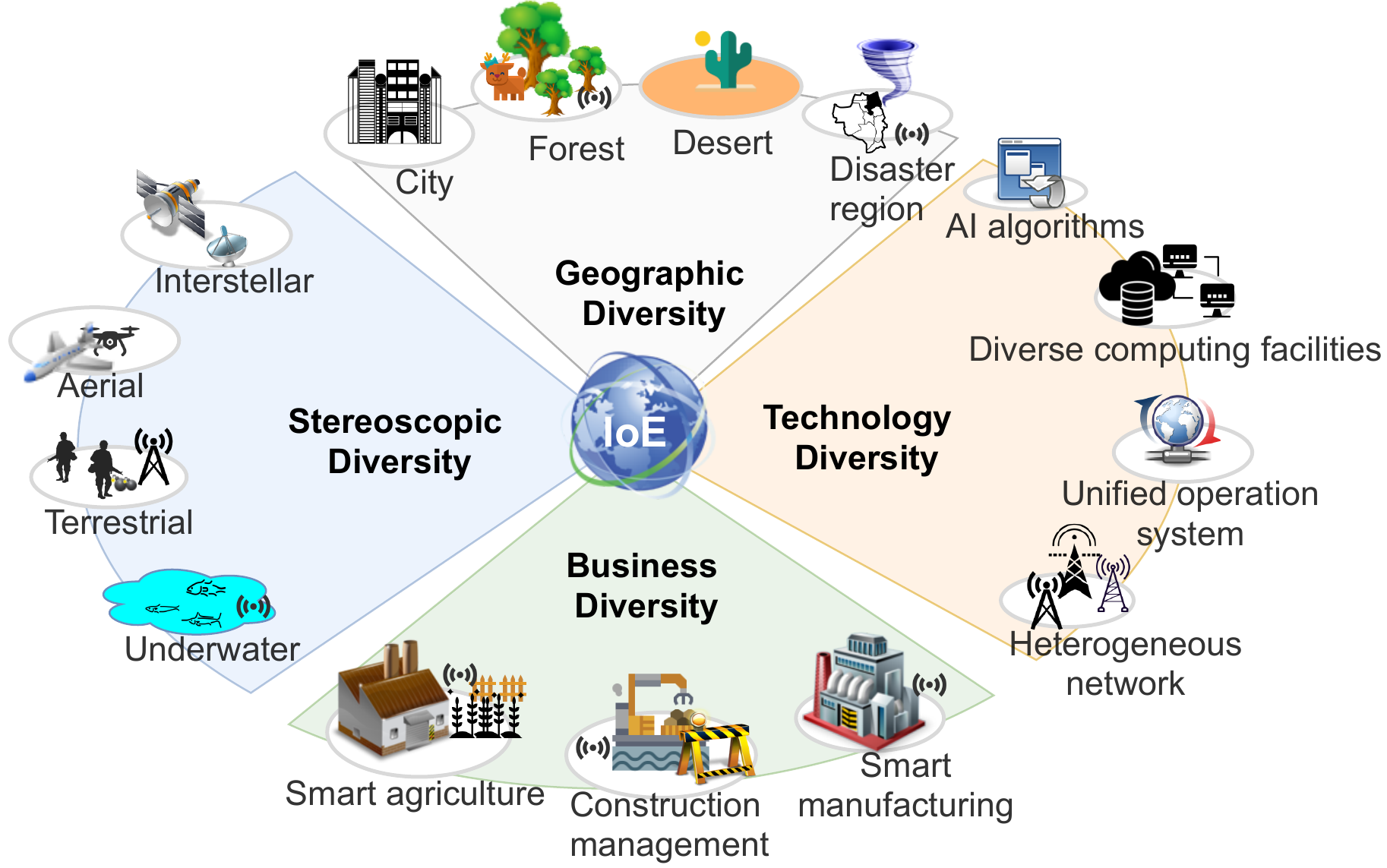}}
		\caption{Enabling applications of IoE's diversity}
		\label{fig: IOEdiverse}
	\end{figure}
	
	On top of distributed computing facilities, big data processing schemes and intelligent algorithms can be executed to enable \textit{everything} intelligence. In particular, big data analytics on IoE data can process massive IoE data and conduct descriptive, diagnostic and predictive analysis \cite{RN112,dai2019manu}. Intelligent algorithms can further extract valuable information on big data so as to make predictive and prescriptive decisions \cite{RN16,fang2019artificial}. Depending on different computing or storage capabilities of cloud servers, edge servers and local nodes, different big data processing and intelligent algorithms can be executed. Strong computing facilities at the cloud side can support the intensive big data processing tasks and intelligent algorithms, e.g., training deep learning models at GPU clusters, so as to enable \emph{cloud intelligence} of IoE. In contrast, edge servers can only support less computing-intensive tasks or intelligent algorithms with less computational complexity (e.g., lightweight or portable deep learning models) \cite{jhzhou:ACM-TIST2019}, thereby enabling edge intelligence~\cite{chen2019improving}. Similarly, local nodes that can only collect and preprocess IoE data are bestowed on local intelligence. 
	
	Big data technologies on top of distributed computing facilities establish a bridge to connect \emph{computing} and \emph{intelligence}. Big data technologies include a series operations such as data acquisition, preprocessing, storage, and preliminary analysis like statistic analysis and data visualization. The heterogeneous IoE data may contain errors, duplicates and redundant values. Therefore, data preprocessing schemes such as data cleaning, compression and integration need to be applied during the data acquisition. In addition, distributed or parallel computing paradigms such as Hadoop, MapReduce and Sparks have also been integrated into big data technologies. Moreover, some preliminary analytical schemes such as descriptive analysis (e.g., statistic analysis and data visualization) and diagnostic analysis are also applied in big data technologies. Recent papers reviewed the usage of big data analytics and intelligent algorithms in IoT scenarios \cite{ur2019role,amanullah2020deep,RN16,RN112}. For example, the authors in \cite{RN16} reviewed big data in IoT field from a historical perspective, covering ubiquitous and pervasive computing, ambient intelligence, and wireless sensor networks. Additionally, the authors in \cite{RN112} reviewed the state-of-the-art studies toward big IoT data analytics. Big IoT data analytics, methods, and technologies for big data mining are discussed. All these papers provide a guideline to better use the merits of big data and intelligent algorithms for IoE's intelligence. 
	
	Intelligent algorithms include both conventional optimization schemes and artificial intelligence (AI) algorithms. Conventional optimization schemes including operational research (OR) programming \cite{fethi2010assessing} and dynamic optimization \cite{mavrovouniotis2017survey} have been well investigated for many years especially in industrial environment, e.g., mechanical automation in production manufacturing \cite{kusiak1987artificial}. AI algorithms \cite{duan2019artificial,zhang20196g} are relatively new to IoE research community. Different from conventional optimization schemes, AI algorithms mainly rely on learning from ambient data and making intelligent decisions. The latest intelligent algorithms including machine learning (ML) \cite{kibria2018big}, deep learning (DL) \cite{buduma2017fundamentals}, reinforcement learning (RL)~\cite{szepesvari2010algorithms}, computer vision (CV), recommendation system (RS), knowledge graph (KG) and collective intelligence (CI, similar to swarm intelligence)~\cite{karaboga2009survey}. These intelligent algorithms outperform statistical methods in diverse tasks like regression, classification, clustering and decision-making. However, different intelligent algorithms have different computing/storage requirements on underlining computing facilities. For example, deep learning models (multi-layer convolutional neural networks) may require extensive training at cutting-edge computing facilities such as GPU servers, which may not be feasible at edge servers and local nodes. Therefore, lightweight or portable intelligent algorithms~\cite{punithavathi2019lightweight,tinyml2020} which can be executed at edge servers or local nodes are expected to be further explored in the future. 
	\subsubsection{Enabling diversity}
	Future IoE devices can be used in diverse fields, including digital sensors for data acquisition and mobile intelligent devices for automated services. As discussed in Section \ref{sec:intro}, we have listed four diversity categories of IoE: geographical diversity, stereoscopic diversity, business diversity, and technology diversity, as shown in Fig. \ref{fig: IOEdiverse}. Specifically, we summarize related applications in every diversity as follows. 
	
	Geographical diversity indicates that IoE applications can be applied in different kinds of geographical regions, including urban, suburban, rural, forests, oceans, and deserts \cite{hashem2016role,lin2018new}. On the other hand, stereoscopic diversity aims at extending the IoE's application range to diverse stereoscopic levels that include the terrestrial, the aerial, underwater and even space \cite{RN114,lin2018new}. In addition, business diversity is to focus on IoE-enabled intelligent business sectors, such as intelligent agriculture, smart manufacturing, smart grid and smart city \cite{dai2019manu,kim2017smart,perera2019unmanned}. Moreover, technology diversity replies on a variety of enabling technologies in IoE. These diverse technologies include different technical aspects such as embedded devices, sensing technologies, communication networks, computing technologies, data processing algorithms, and AI algorithms. 
	
	It is an inevitable trend that future IoE applications become the fusion of different diversities. This fusion has already been discussed in previous literature. For example, the authors in  \cite{RN17} explored the integration of heterogeneous networks of IoT. The authors in \cite{zarca2019security} investigated the global virtual computing system combining cloud, edge, fog, and local. The authors in \cite{RN109} have concentrated on the design of the unified operating system for IoT. Practically, the aforementioned four IoE diversities can benefit from each other. The most obvious fact is that the application in different diversities can coexist with each other for a common suitable reason. For example, urban services are more suitable for terrestrial and business applications. However, forests and deserts scenarios require more aerial and even interstellar networks for flexible and on-demand IoE services. Therefore, different diversities can always fulfill the corresponding application demands with each other. As a result, one open issue in future IoE is to coordinate different applications in the same diversity category or to orchestrate the same application across different diversity categories.

	\subsection{Challenges of IoE}
	\label{subsec:challenges}
\begin{figure}[t]
		\centerline{\includegraphics[width=\columnwidth]{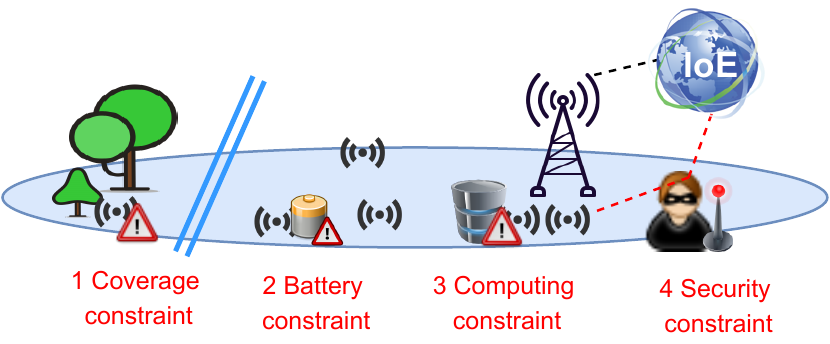}}
		\caption{Challenges of IoE}
		\label{fig: ioecha}
	\end{figure}

	Although the aforementioned enabling technologies can potentially realize IoE's scalability, intelligence, and diversity, a number of challenges are rising when implementing those enabling technologies. Attributed to the restricted resources (e.g., network infrastructure, spectrum access, hardware cost), the challenges of IoE are reflected in the following four constraints: coverage constraint, battery (energy) constraint, computing constraint, and security constraint, as shown in Fig. \ref{fig: ioecha}. Next, we will discuss the four constraints in detail.
	
	\begin{enumerate}[1)]
		\item \textit{Coverage constraint:} It is difficult (or expensive) to deploy IoE communication infrastructures at harsh and rural geographical areas. As a result, the IoE nodes in these areas are hard to be covered. We name such challenge as \emph{coverage constraint} and such areas as \emph{coverage-constrained} areas. Deploying the incumbent communication networks in the coverage-constrained areas is not budget-friendly due to the imbalance between costs and benefits in network construction. A key observation is that the IoE projects in cover-constrained areas do not have such strong communication demands at \textit{anytime} from \textit{anywhere}. Therefore, a flexible and recoverable coverage is the most cost-effective solution to support on-demand IoE communications in the coverage-constrained areas for a particular period. 
		\item \textit{Battery constraint:} IoE nodes suffer from the battery constraint. IoE nodes are generally power-limited due to either hardware cost or portability concerns. In particular, IoE tends to use low power or even battery-free communication technologies to access network infrastructure nodes (e.g., BS, AP and IoT gateways). One inevitable fact is that battery-limited nodes are easily exhausted and eventually lose connections with IoE. The connection-lost problem becomes even severe in the coverage-constrained areas. Hence, it is necessary to develop sustainable energy supplying and the recovery mechanisms for battery-constrained nodes.
		
		\item \textit{Computing constraint:} Most of terminal nodes of IoE do not have enough computing capability to process local intelligent algorithms. We call these nodes as computing-constrained nodes. The traditional solutions for IoE are to transmit all data to remote cloud servers which can offer a centralized intelligence for big data processing. However, the cloud computing paradigm also brings a high latency, which cannot be ignored especially for the future latency-sensitive IoE applications. Meanwhile, the increasing number of IoE computing tasks will not only cause the burden to the cloud servers but also lead to the congestion at backbone communications of IoE as well as privacy-leakage risks. To alleviate the bottlenecks at cloud servers, it is expected to fully utilize both edge and local computing resources as a supplementary of enabling everything intelligence. 	
		\item \textit{Security constraint:} Many potential security risks are encountered in IoE, attributed to the vulnerabilities of communication protocols as well as resource limitations of IoE nodes. In particular, the current IoE mostly adopts the low-cost and simplified access protocols (i.e., NB-IoT, LoWPAN) in order to reduce network cost while it makes the communications be vulnerable to malicious attacks such as eavesdropping and forging. On the one hand, the data emitted from end nodes can be wiretapped (or eavesdropped) by malicious nodes; on the other hand, and pseudo-base-stations can easily forge the normal IoE communication links to obtain IoE data \cite{khattak2019perception}. Therefore, an effective but easy-deployed security mechanism is required to protect IoE communications from malicious attacks.
	\end{enumerate}	
	
	\textbf{Discussion:} To overcome these challenges, we need to take some effective countermeasures: i) building flexible and recoverable networks to extend IoE's coverage; ii) developing sustainable energy-supply mechanisms to prolong the life-cycle of IoE's nodes; iii) orchestrating the edge computing with local and cloud computing to optimize various computing resources and schedule diverse computing tasks; iv) designing reliable security solutions to protect the data in the ubiquitous IoE from malicious attacks. Particularly, UAVs have enormous potentials to provide an attractive solution to address four challenges of IoE, thanks for their flexibility and on-demand deployment manner. Combining with existing communication networks, wireless power transfer technologies, edge computing, and physical jamming, UAVs can build an extended network of IoE with sustainable power, edge intelligence, and physical security protection. 	
	
	\section{Overview of UAV}
	\label{sec:uav}
	
	\begin{figure*}[t]
		\centerline{\includegraphics[width=15cm]{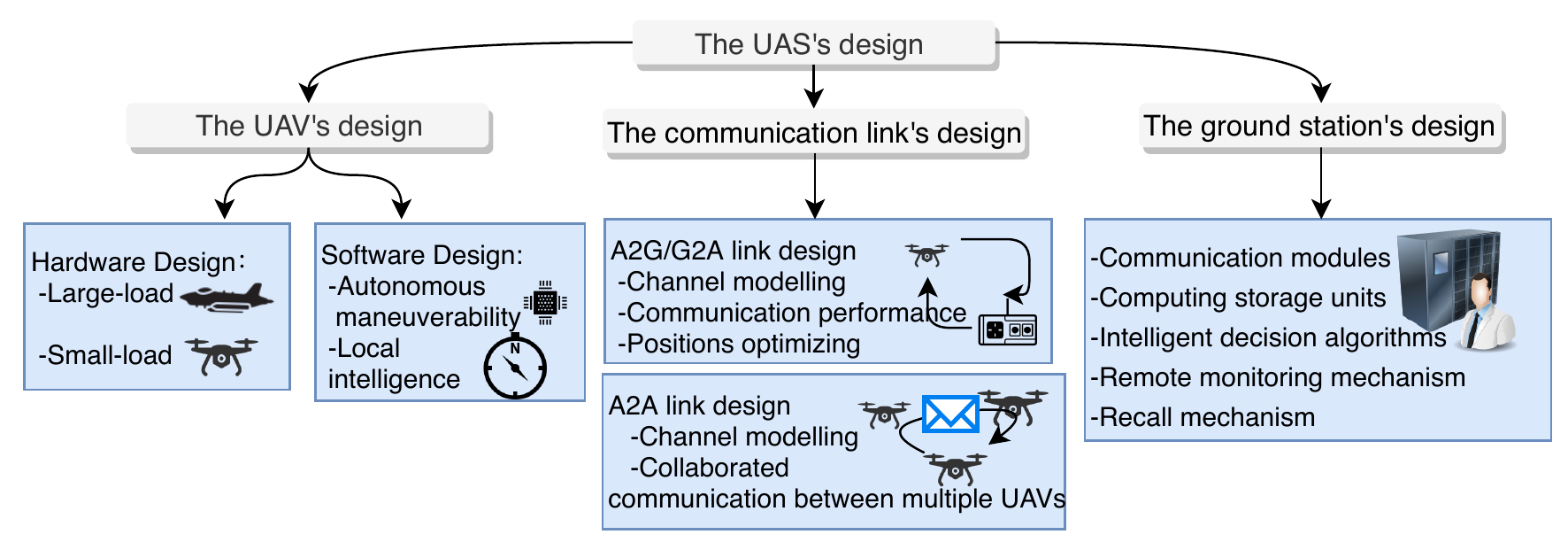}}
		\caption{Design method of the unmanned aircraft system (UAS)}
		\label{fig: uas}
	\end{figure*}
	
	In recent decades, UAVs have attracted a significant attention from both academia and industry due to their boundless services. Such a wide variety of UAV applications rely on controllable maneuverability technologies such as flight trajectory optimization and obstacles avoiding. UAVs' maneuverability can realized by on-board intelligent algorithms and remote control from ground pilots (or ground control stations). This section first introduces the unmanned aircraft system (UAS) in Section~\ref{subsec:UAS} and UAV communication networks in Section~\ref{subsec: uavcommunication} to demonstrate UAVs' controllable maneuverability.
	
	\subsection{Unmanned aircraft system (UAS)}
	\label{subsec:UAS}
	
	UAS provides the cooperated control services for UAV applications. The control services include flight control, information processing, tasks scheduling, etc. A typical UAS is composed of three components: UAVs, the ground-based controller/control station, and the communication links between them \cite{gupta2013review,manitta2016unmanned,chan2018reconfigurable}. UAS is crucial to achieve UAVs' flight control and their task scheduling. For large UAVs, the take-off and the landing are controlled by ground control stations. After the UAV reaches the cruising altitude, the automatic driving mode is switched on and the corresponding flight task begins. When a UAV performs autonomous flight control and on-board task scheduling, the function of the ground controller may be integrated on-board, thus the autonomous UAV itself becomes a UAS. In contrast, the small UAV mainly relies on its small-and-nimble superiority in conducting flight tasks with strong maneuverability. The small size makes it impossible to carry many mission equipment or large on-board autonomous system. As a result, the small UAVs tend to depend on the remote control from ground stations. In this regard, their flight range is limited by the communication range between the UAV and the ground station. 
	
	It is necessary to design a dedicated UAS for specific applications such as video surveillance, automatic localization and tracking. As shown in Fig. \ref{fig: uas}, the design of a dedicated UAS includes three aspects, i.e., the UAV's design, the ground control station's design, and the communication link's design. We now discuss these design aspects as follows.
	
	\subsubsection{The UAV's design}
	\label{designofuav}
	UAV that can conduct flexible flying tasks, was early used for military unmanned air-crafts that own a variety of mission equipment \cite{RN19,RN106,RN96,RN44,RN105}. In order to perform a variety of unmanned military missions (e.g., reconnaissance, tempting enemy and target striking), for a long time, military UAVs usually adopt the structure with the fixed-wings to carry large loads. In addition, small UAVs are designed with short battery life to carry small loads. Therefore, they have mainly been developed for civilian applications. One example is using small UAVs for aerial photography in a particular region. This activity has become very popular among aerial enthusiasts. Compared with a large fixed-wing UAV, a small rotary-wing UAV has more flexibility. After being special configured, either large UAVs or small UAVs can be applied for various commercial purposes, such as air logistics, emergency communications, and airbase stations \cite{RN28,8600892}.  
	Next, we will introduce UAVs' design via two aspects: hardware design and software design \cite{DJIdrone1,DJIdrone2,RN86}. 
	\begin{enumerate}[$\bullet$]
		\item \textit{Hardware design:} UAVs' hardware design focus on the hardware configuration for suitable flight mechanics (i.e., flying wings and engine) and mission equipment (i.e., the communication module, computation chip, and other carrying equipment). UAVs' load-size has large impacts on their hardware design in specific flight tasks. 
		For \textit{long-time and long-distance flight tasks} (such as air-to-sea image telemetry or air-to-ground forest fire early warning), stability and persistence are main objectives for hardware design of UAVs. In this regard, the large-size UAVs are suitable to perform flight tasks requiring high stability and persistence. Since the large-size UAVs can be configured with persistent flight functions (including fixed-wings, long-term battery, and stable engines) and strong mission ability (including on-board cameras, signal processing, carrying capability). For \textit{limited-range and high-maneuverability flight tasks} (such as urban criminal tracking, instant sampling of accident scenes), mobility and efficiency are two main factors impacting the hardware design. The small-size UAVs are suitable to perform missions requiring high mobility and efficiency. To achieve this goal, they will be configured with flexible flight structures (e.g., rotor-wings, limited battery capacity) and light mission equipment (containing communication module and on-board cameras). 	
		\item \textit{Software design:} UAV's software design aims to develop dedicated on-board algorithms for autonomous flight decision-making. By collecting the ambient data, processing and analyzing the data, UAVs can make flight decisions of autonomous maneuverability and achieve local intelligence. Autonomous maneuverability requires dedicated designs for stable flight in complex geological/meteorological environments. Basic maneuvering algorithms include obstacle detection, collision avoidance, motion adjusting, and trajectory planning. Local intelligence is necessary since offline mission algorithms for aerial delivery, emergency communication, geological surveys need to be implemented at UAVs. For instance, UAVs can be configured with the dedicated communication protocols and planned with a given trajectory to perform aerial delivery \cite{8600892}, or to conduct the remote data collection \cite{RN28}. 
	\end{enumerate}
	
	\subsubsection{The communication link's design} There are two kinds of wireless communication links in UAS. One link is the Air-to-Ground (A2G)/Ground-to-Air (G2A) link that connects the UAV with the ground node. Another link is the Air-to-Air (A2A) link that connects multiple UAVs together when they conduct collaborative multi-UAV flight tasks. Both two kinds of links require the specified design to work in particular communication scenarios. The design principles include communication connectivity, the fly trajectory of UAVs, and the successful probability of flight tasks. Corresponding studies investigating two kinds of links are summarized as follows.
	\begin{enumerate}[$\bullet$]
		\item The studies on A2G/G2A communications aim at improving the communication performance by optimizing locations of UAVs. Generally, the A2G link considers optimizing the communication of feedback information from UAVs to ground stations. The G2A link considers the communication of controlling signals from ground stations to UAVs. To optimize the A2G/G2A communications, locations of UAVs are in dedicated selection to satisfy the link quality. This analysis process can usually be accomplished by firstly building a reasonable channel gain model of A2G/G2A, then setting the condition of communication performance (related to channel gain), and finally optimizing flying locations of UAVs subjected to the condition. The above investigation steps are contained in the related work on UAV-aided communication networks. The corresponding studies are summarized as shown in Section \ref{subsec: uavcommunication}.
		\item Different from the investigation of A2G/G2A, the studies investigating A2A aim at accomplishing the collaborative flight tasks by multiple UAVs. The A2A link establishes communications between multiple UAVs. In this regard, the design objective of A2A is to ensure collaborative communication performance in a muti-UAV network. Additionally, A2A can be easily modeled as a line-of-sight (LoS) propagation, and the link quality of A2A is susceptible with the mobility of multiple UAVs. Hence the investigation direction is transformed from improving the A2A link quality to the collaborated position scheduling of multiple UAVs. Due to UAVs' mobility from time to time, the collaborated positions are essentially varied, which further leads to the necessity of routing protocols. The particular design in A2A links has been covered in the related work on multi-UAV Ad Hoc networks. The corresponding studies are summarized as shown in Section \ref{subsec: uavcommunication}.
	\end{enumerate}	
	
	\begin{table*}[t]
		\centering
		\caption{UAV communication networks}
		\renewcommand{\arraystretch}{1.75}
		\label{tab: uavcom}
		\begin{tabular}{|p{2.5cm}|p{6cm}|p{6cm}|}
			\hline
			\textbf{Introduction} & \textbf{Multi-UAV Ad Hoc networks}& \textbf{UAV-aided communication networks}\\ \hline\hline
			\textit{Architecture} & Self-organized networks	& Edge side of current networks
			\\ \hline
			\textit{Characteristics}	&Flexible topology	&Flexible access
			\\ \hline
			\multirow{2}*{\textit{Applications}}&Collaborative communications \cite{RN22,RN71,UAVdata2018}&Relay communications \cite{alzenad20173,lyu2016placement,zhang2017joint}\\
			&Emergency communications \cite{RN42}&Flying base stations \cite{Choi2014UAV,semsch2009autonomous,motlagh2017uav}\\ \hline	
			\textit{Wireless links}&A2A links&A2G/G2A links	
			\\ \hline
			\textit{Research issues}&Routing design	\cite{wu2019fundamental,RN19,RN18,RN61,sathyaraj2008multiple}&Trajectory optimization \cite{net2016trajectory,net2016trajectory}\\ \hline
			\multirow{4}*{\textit{Research methods}}&Global throughput maximization \cite{wu2019minimum,RN48,xu2018uav}&G2A/A2G throughput maximization \cite{zhang2017joint,8760267,8737778}\\
			&Maximizing energy efficiency \cite{hua2019energy}& Maximizing energy efficiency \cite{RN56,8758340}\\
			&Dynamic topology/swarm \cite{kow2019quav,8750806}&A2G/G2A localization \cite{8811579,8422460}\\	
			&Multi-UAV's resource allocation \cite{RN53,8758997,8727504}&Edge computing \cite{8607062}\\
			&Dense UAV networks \cite{8473486}&Ground node access \cite{8641428,8485372}\\
			\hline	
		\end{tabular}
	\end{table*}
	
	\subsubsection{The ground control station's design}
	The ground control station with the responsibilities for UAV's tasks scheduling and remote communications, is the decision center of the entire UAS. To accomplish the responsibilities, a ground control station is designed with the following functions: wireless communication modules, computing/storage units with large enough capacity, intelligent decision algorithms for centralized mission planning, remote monitoring mechanisms, and recall mechanisms. The communication modules are crucial to achieve remote control for UAVs. Hence its design should match with the communication modules of UAVs. 
	The computing/storage units are designed to satisfy the data processing requirements. Intelligent decision algorithms are mainly used to schedule UAV tasks with objectives such as the minimized overall time, optimized trajectory and optimized resources, etc. \cite{dronelife}. The remote monitoring mechanisms are used to globally monitor UAVs' flying states, which can be achieved by periodical receiving the information feedback from UAVs. The recall mechanisms are designed for a controllable recall for abnormal cases (e.g., the cases when flying tasks require interruption, or the cases that UAVs are detected as energy shortage).
	
	\subsection{UAV communication networks}
	\label{subsec: uavcommunication}

	Compared with UAS that offers a control system for a specified UAV mission, UAV communication networks mainly offer stable communications between UAVs or between the UAV and the ground station. The state-of-the-art literature provides two classes of UAV communication networks: multi-UAV Ad Hoc networks and UAV-aided communication networks. Table \ref{tab: uavcom} gives a brief summary of these two UAV communication networks. More details are introduced as follows.
	\subsubsection{Multi-UAV Ad Hoc networks}
	Multi-UAV Ad Hoc networks are self-organized with high autonomy, and are usually independent of incumbent mobile networks. Multi-UAV Ad Hoc networks can support mobile connections and coverage in some specified occasions such as emergency communication networks, aerial surveillance fleet, aerial sensor networks. For instance, the authors in \cite{RN22} presented such a typical instance, in which a multiple-UAV network assists the vehicular-to-vehicular (V2V) communications in the regions with poor connectivity to infrastructure-based networks. Besides, multi-UAV networks can perform the data acquisition tasks in large-scale sensor networks \cite{RN42,RN71,UAVdata2018}.

	The studies on multi-UAV networks mainly focus on performance improvement of A2A links between multiple UAVs. Particularly, the A2A performance includes wireless connectivity and information interaction. Due to the mobility of multiple UAVs, the channel quality of A2A links is unstable, consequently leading to the dynamic network topology. The related studies include two directions: multi-UAV's trajectory optimization and dynamic routing protocol. The two study directions have gained significant attention for multi-UAV networks, as shown in several surveys \cite{RN19,RN18,RN61}. For instance, \cite{RN19} gives a comprehensive summary of routing protocols for multi-UAV communication networks, including self-organization, disruption tolerance, SDN control, seamless handover, and energy efficiency. Meanwhile, \cite{RN18} compares the performance of existing routing protocols that are classified into two categories: single-hop routing and multi-hop routing, where the comparison metrics include load balancing, loop-free ability, route update method, dynamic robustness, energy efficiency, and so on.

	In addition, previous studies investigated the trajectory optimization of multi-UAV networks when a specified routing scheme, i.e., single-hop or multiple-hop scheme is selected. For instance, the authors in \cite{RN48,xu2018uav} investigated the trajectory optimization joint with power control. Ref. \cite{RN48,xu2018uav} considered a single-hop routing scheme, in which every ground node only accesses one UAV and multiple UAVs cooperatively serve for a groups of ground nodes. Additionally, the authors in \cite{RN53,sathyaraj2008multiple,net2016trajectory} investigated the trajectory optimization for the case of a multi-hop routing scheme, in which multiple UAVs provided collaborative communications to a group of ground nodes, with the objective of end-to-end throughput maximization.	
	
	\begin{figure*}[t]
		\centering
		\subfigure[a][Ubiquitous connections]{\includegraphics[width=8cm]{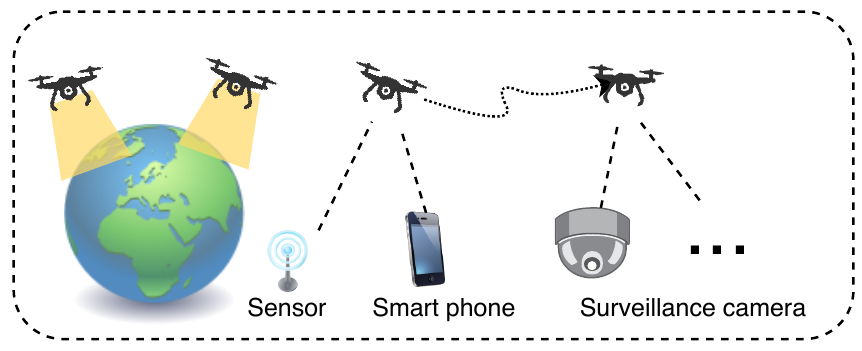}}
		\subfigure[b][Aerial intelligence]{\includegraphics[width=8cm]{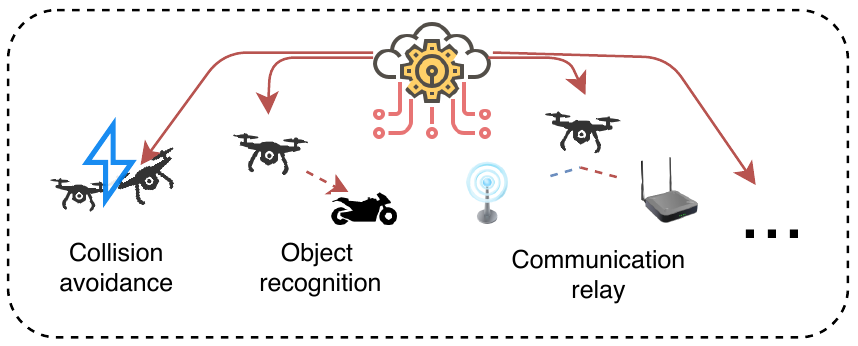}}
		\subfigure[c][Self-maintenance of communications]{\includegraphics[width=8cm]{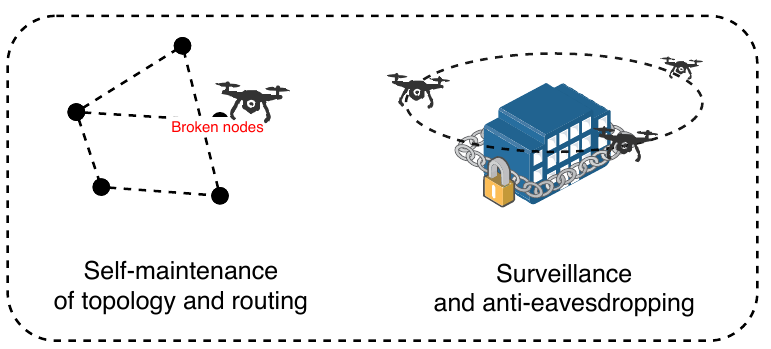}}
		\subfigure[d][Sensor powering and deployment]{\includegraphics[width=8cm]{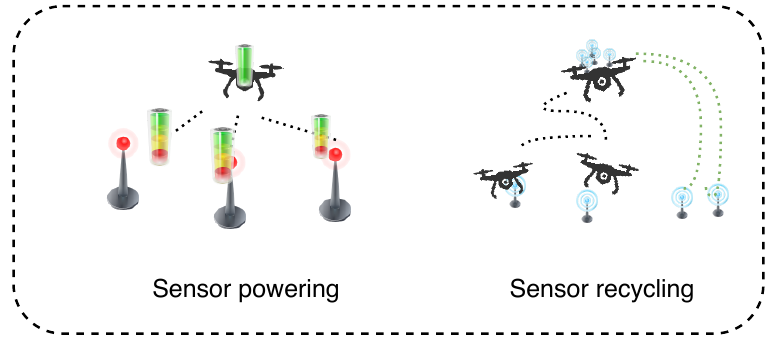}}
		\caption{Opportunities of UAV in IoE}
		\label{fig: opportunities}
	\end{figure*}
	
	\subsubsection{UAV-aided communication networks}	
	\label{subsubsec: uavedge}
	The UAV-aided communication network is the extended edge network of existing networks. Different from independent self-organized architecture of multi-UAV ad hoc networks, UAV-aided communication networks are infrastructure-based topology, depending on the existing network access points. Therefore, UAV-aided communication networks adopt the wireless protocols of the aided networks (e.g., MCNs, and WLANs). 
	
	In a UAV-aided communication network, UAVs can be deployed as edge nodes with network functions such as flying base-stations (e.g., \cite{alzenad20173,lyu2016placement}), relay nodes \cite{zhang2017joint,Choi2014UAV}, or terminal nodes (e.g., the aerial surveillance camera) \cite{semsch2009autonomous,motlagh2017uav}. The objective of UAV-aided communication networks is to provide flexible communication services in areas losing connections with incumbent networks. Herein, UAVs play a role of relay/terminal nodes which connect with the disconnected nodes thereby establish the communication links. However, practical scenarios are faced with variable factors, such as uncertain locations of the disconnected nodes, and diverse communications protocols. All these undetermined factors pose the challenges in optimizing the flying trajectory of UAVs. The optimization requires the dedicated design for specific scenarios. Taking UAV-enabled patrolling as an example, in which the flying trajectory of UAVs is required to satisfy the mobile connectivity, i.e., mobile A2G/G2A links at UAV-aided communication networks. In this example, the objective of UAVs' trajectory optimization is to ensure the coverage for the task region while keeping the connections with the network.
	
	For UAV-aided communication networks, most of studies are mainly related to trajectory optimization of UAVs. In previous studies, trajectory optimization was done by optimizing various performance metrics including QoS of communication links, sufficient coverage \cite{net2016trajectory}, time-efficiency, energy efficiency \cite{RN56}, outage probability, etc. For instance, the authors in \cite{de2010uav} presented UAVs' trajectory optimization through an iterative algorithm that reduces the disconnected nodes for covering multiple isolated WSN nodes. Similarly, \cite{alzenad20173} proposed an optimal placement algorithm for UAV-base stations by maximizing the number of covered users. In addition, \cite{Choi2014UAV} designed the trajectory of a single relay UAV by maximizing energy-efficiency. Moreover, \cite{zhang2017joint} optimized UAVs' trajectory by minimizing the outage probability, where UAVs worked as an amplify-and-forward relay.

	\textbf{Summary:} The studies on UAVs mainly concentrate on two critical issues: one is to investigate the UAS for controllable maneuverability of UAVs, and the other is to study the UAV communication networks to ensure link connectivity (including the connectivity of A2A links and A2G/G2A links). These two issues share a common research problem - UAV's mobility design (i.e., trajectory optimization). According to recent studies on  UAVs \cite{RN54,RN56,RN53,RN48}, UAVs bring opportunities to enable IoE. In next section, we will discuss the convergence of UAVs and IoE. 
	
	
	\section{Convergence of UAV and IoE}
	\label{sec:uavioe}
	As discussed in Section \ref{sec:uav}, UAVs can be employed to serve as aerial base stations, data collectors, jammers, monitors, edge computing servers, power suppliers, reclaimers for IoE. Accordingly, UAVs offer a solution to fulfill three expectation of IoE with provision of extended coverage, flexible intelligence, and diverse applications. In this section, we first investigate the opportunities for applying UAVs in IoE in Section~\ref{subsec:opporueioe}.  We then present a solution namely Ue-IoE to integrate UAV technologies with IoE in Section~\ref{subsec:Ue-IoE}. 
	
	\subsection{Opportunities brought by UAVs}
	\label{subsec:opporueioe}
	With high mobility and reconfigurability, UAVs can potentially address the four constraints of IoE. UAVs bring four opportunities to IoE, i.e., ubiquitous connections, aerial intelligence, self-maintenance of communications, sensor powering and deployment, as shown in Fig. \ref{fig: opportunities}. 
	\subsubsection{Ubiquitous connections}
	\label{subsubsec:opp1}
	Ubiquitous connections are the necessity for IoE to cover \textit{everywhere} and connect \textit{everything}. Flexible UAV communication networks can help IoE to extend its coverage, thereby achieving ubiquitous connection. As discussed in Section \ref{subsubsec:sclableueioe}, UAVs can extend the communication network to the areas with weak-connection via UAV-aided communications networks; UAVs can increase coverage to the areas without network infrastructures through multi-UAV Ad Hoc networks. Accordingly, ubiquitous connections can be achieved. 
	\subsubsection{Aerial intelligence}
	\label{subsubsec:opp3}
	UAVs can enable \emph{aerial intelligence} by collecting the surround data and then performing on-board intelligent algorithms. The surround data are collected from either the UAV itself or a cluster of ambient sensors. The intelligent algorithms are dedicated to fulfill different requirements, such as autonomous collision avoidance, adaptive flight gesture adjustment, trajectory optimization for data collection. Thus, UAVs can support many aerial intelligent applications. A typical application is tracking a moving target by UAVs \cite{RN52}. Meanwhile, UAVs can play as an aerial command-maker to conduct intelligent perception and make decision for computing-constrained IoE nodes.

	\subsubsection{Self-maintenance of communications}
	\label{subsubsec:opp4}
	UAVs can support self-maintenance for IoE communications. Due to various unstable factors from either the urban or natural environments, some IoE nodes are easily destroyed and even lost. Thus, IoE faces the risk of losing connections. To address this issue, UAVs can be dispatched to redeploy IoE nodes so as to restore the lost links. In addition, some IoE communications are susceptible to malicious attacks such as eavesdropping and forging attacks. These risks may cause huge economic losses when they occur in high-confidential communications since they may cause malfunction in smart manufacturing or in intelligent transport system. In this regard, UAVs can be employed as friendly-jammers to form a protective barrier for IoE's physical communication of IoE. The authors in \cite{wang2019novel} present a UAV-based friendly jamming scheme to interfere with eavesdropper's communications. In addition, the authors in \cite{kharchenko2018cybersecurity,he2018flight} offer enhanced encryption security by configuring the dedicated protocols in UAVs. 
	\subsubsection{Sensor powering and recycling}
	\label{subsubsec:opp2}
	UAVs can be re-configured to perform sensor powering and recycling tasks, thus can improve sustainability of IoE. The resource-constrained IoE nodes can usually easily run out of battery and then get discarded. This phenomenon not only results in a big waste but also causes high pollution. UAVs can potentially avoid this waste or pollution by charging IoE nodes with energy. In particular, UAVs can charge these nodes by wireless power transfer (WPT) technologies \cite{RN54} and even can achieve simultaneous wireless information and power transfer (SWIPT) \cite{RN64}. In addition, UAVs can recycle the damaged nodes or sensors with disabled function or place them with new nodes.
	
	\subsection{UAV-enabled IoE}
	\label{subsec:Ue-IoE}
	Motivated the opportunities brought by UAVs to IoE, we can comprehensively apply UAVs to enhance IoE's capabilities. We present a UAV-enabled IoE solution (namely Ue-IoE in short in the rest of this paper) to fulfill three expectations of IoE. In particular, we will introduce Ue-IoE based on three solutions to the three expectations, i.e., UAV-enabled scalability to IoE, UAV-enabled intelligence to IoE, and UAV-enabled diversity to IoE. 
	\subsubsection{UAV-enabled scalability to IoE}
	\label{subsubsec:sclableueioe}
	Cooperating with existing networks (i.e., WLAN, MCN, LPWAN, satellite network), UAVs can enable a scalable IoE to cover \textit{everywhere} and connect \textit{everything}. UAVs can maximize the coverage of IoE by extending the current IoE network to two main kinds of areas: areas with weak-connection, areas without network infrastructures.
	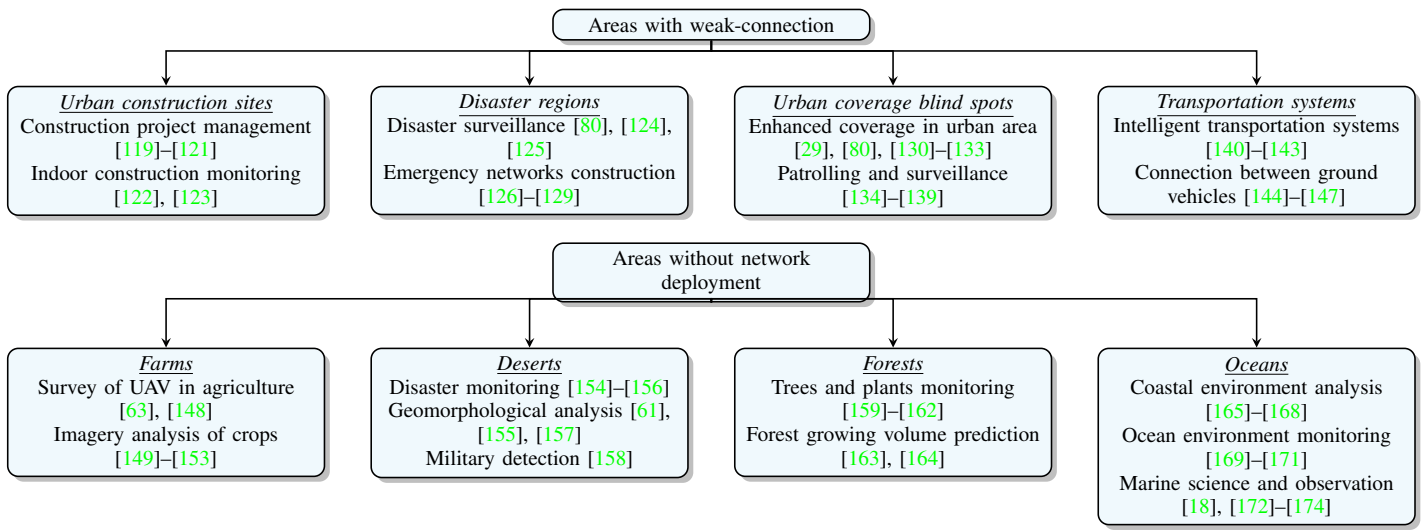
\begin{figure*}[t]
		\centering
		\small
		\subfigure{\begin{forest}
				for tree={              
					draw, semithick, rounded corners,
					text width = 40mm, text badly centered,
					edge = {draw, semithick, -stealth},
					anchor = north,
					grow = south,
					forked edge,            
					s sep = 6mm,    
					l sep = 6mm,    
					fork sep = 3.5mm,  
					tier/.option=level,
					fill=cyan!5,
				}
				[Areas with weak-connection
				[\underline{\textit{Urban construction sites}}\\	Construction project management \cite{zhou2018multidimensional,li2018applications,hubbard2015feasibility} \\ 
				Indoor construction monitoring \cite{hamledari2016inpro,moud2018current}]
				[\underline{\textit{Disaster regions}} \\
				Disaster surveillance \cite{abdallah2019efficient,RN71,maza2011experimental} \\ 
				Emergency networks construction \cite{zhao2019uav,tuna2014unmanned,deruyck2018designing,ibrah2018optimization}
				]
				[\underline{\textit{Urban coverage blind spots}} \\
				Enhanced coverage in urban area \cite{sujit2015decentralized,RN71,orsino2017effects,kim2018designing,naqvi2018drone,andreev2019future} \\ 
				Patrolling and surveillance \cite{RN35,wu2018novel,wang2016optimal,seok2017unpredictably,zhou2018continuous,chen2015decentralized}
				]
				[\underline{\textit{Transportation systems}} \\
				Intelligent transportation systems \cite{menouar2017uav,8307444,bertrand2018evaluating,reshma2016security} \\ 
				Connection between ground vehicles \cite{oubbati2019uav,zhou2015multi,hadiwardoyo2018evaluating,hadiwardoyo2018experimental}
				]
				]	
		\end{forest}}
		\subfigure{\begin{forest}
				for tree={              
					draw, semithick, rounded corners,
					text width = 40mm, text badly centered,
					edge = {draw, semithick, -stealth},
					anchor = north,
					grow = south,
					forked edge,            
					s sep = 6mm,    
					l sep = 6mm,    
					fork sep = 3.5mm,  
					tier/.option=level,
					fill=cyan!5,
				}
				[Areas without network deployment
				[\underline{\textit{Farms}}\\ 
				Survey of UAV in agriculture \cite{perera2019unmanned,milics2019application} \\ 
				Imagery analysis of crops \cite{RN97,RN92,RN82,RN100,RN98}
				]
				[\underline{\textit{Deserts}}\\
				Disaster monitoring \cite{abdelkader2013uav,sparelli2018geomorphological,sun2018analysis} \\ 
				Geomorphological analysis \cite{sparelli2018geomorphological,sankey2018uav,lin2018new}\\
				Military detection \cite{abushahma2019comparative} \\
				]
				[\underline{\textit{Forests}}\\
				Trees and plants monitoring \cite{RN102,RN87,RN88,otero2018managing} \\ 
				Forest growing volume prediction \cite{puliti2018combining,giannetti2018new}
				]
				[\underline{\textit{Oceans}} \\
				Coastal environment analysis \cite{RN101,RN120,matsuba2018nearshore,gooday2018assessment} \\ 
				Ocean environment monitoring \cite{wang2019accurate,bousquet2018unav,fontanesi2019over}\\ 
				Marine science and observation \cite{wu2018uav,johnston2019unoccupied,aniceto2018monitoring,rieucau2018using}]
				]	
		\end{forest}}
		\caption{UAV-enabled scalability to IoE}
		\label{fig: coveragearea}
	\end{figure*}
	\begin{enumerate}[(i)]
		\item \textit{Areas with weak-connection.} These areas are usually covered by the existing networks such as WLAN, MCN, and LPWAN while they are always in the weak-connection state due to the complex geographical environment (e.g., obstacles) and harsh environment. Four typical regions are \textit{Construction sites in urban}, \textit{Disaster regions in urban}, \textit{Coverage blind spots in city} and the \textit{Transportation road}. These regions are scattered with various obstructions, resulting in the unstable wireless links between above areas with the existing APs. To address this problem, UAV-aided communication networks can provide a flexible network. In this sense, UAVs can play as the on-demand relay nodes or base stations or gateways to connect the IoE nodes. In addition, A2G or A2A links of UAVs that are typically LoS can avoid the obstacles especially in complex geographical environment. 
		\item \textit{Areas without network infrastructure.} These areas are generally remote and lack of inhabitants. Hence no network infrastructures are deployed in such areas. Four typical regions are \textit{farms, deserts, forests}, and \textit{oceans}. To cover these regions, two communication schemes are required: 1) building an independent network for every specific area; 2) designing the access scheme for this isolated network to connect with existing IoE networks. Multi-UAV Ad Hoc networks can be applied to achieve above two goals. The multi-UAV network cannot only independently cover the remote areas but also execute many specific tasks (including relay communications, remote sensing, data acquisition, etc \cite{RN29}). From a global perspective, multi-UAV Ad Hoc networks are required to enable the specified coverage when UAVs fly to remote areas. In addition, UAV-aided communication networks are required for IoE's access when UAVs fly back to the ground control center.	 
	\end{enumerate} 
	Fig. \ref{fig: coveragearea} shows a tree diagram to categorize the corresponding studies of using UAVs to address the above two issues. 
	For the areas with weak-connection, we list four typical regions that are construction sites in urban, disaster regions in urban, blind coverage spots in the city, and the transportation road. In these areas, some recent studies use UAVs to offer an extended network coverage and perform some specified applications such as construction project management, emergency networks, patrolling, and intelligent transportation. Moreover, for the areas without network deployment, UAVs are strongly in demand for achieving a series of unmanned and remote applications in four typical regions (i.e., farms, deserts, forest, and ocean). Some particular applications in these four regions are smart farms, disaster monitoring in deserts, plants monitoring in forests, etc.    
	\subsubsection{UAV-enabled intelligence to IoE} 
	\label{subsubsec:intellueioe}
	UAVs can enable IoE's intelligence by bestowing lightweight AI algorithms at UAVs. In particular, UAVs can make smart decisions or controls, thereby further enabling intelligence to themselves as well as IoE nodes. Next, we introduce the UAV-enabled intelligence of IoE via two respects: UAV-enabled intelligent networks and UAV-enabled intelligent aerial services.
	\begin{figure*}[t]
		\centering
		\small
		\subfigure{\begin{forest}
				for tree={              
					draw, semithick, rounded corners,
					text width = 50mm, text badly centered,
					edge = {draw, semithick, -stealth},
					anchor = north,
					grow = south,
					forked edge,            
					s sep = 6mm,    
					l sep = 6mm,    
					fork sep = 3.5mm,  
					tier/.option=level,
					fill=magenta!5,
				}
				[UAV-enabled intelligent networks
				[\underline{\textit{Intelligent communication layer}}\\Optimizing the position and trajectory \cite{deruyck2018designing,zhao2019uav,oubbati2019uav,naqvi2018drone,cao2018mobile}]
				[\underline{\textit{Intelligent network layer}}\\ Routing selection \cite{RN61,RN18,RN19} \\Congestion control \cite{barritt2017operating,ur2017deployment,yuan2016software,kumar2017sdn,zhang2018sdn}\\Smart access \cite{bai2019energy,RN129}
				]
				[\underline{\textit{Intelligent application layer}}\\High-level control \cite{khosiawan2016system,madni2018formal,luo2015uav,saigh2017uav}
				]
				]	
		\end{forest}}
		\subfigure{\begin{forest}
				for tree={              
					draw, semithick, rounded corners,
					text width = 55mm, text badly centered,
					edge = {draw, semithick, -stealth},
					anchor = north,
					grow = south,
					forked edge,            
					s sep = 6mm,    
					l sep = 6mm,    
					fork sep = 3.5mm,  
					tier/.option=level,
					fill=magenta!5,
				}
				[UAV-enabled intelligent aerial services
				[\underline{\textit{Intelligent local computing services}} \\Zero transmission delay \\ Lightweight intelligence \\ \cite{RN107,aasen2015generating,van2018real,eling2015real}\\ ]
				[\underline{\textit{Intelligent edge computing services}}\\ Less transmission delay \\ Mediate intelligence \\ \cite{luo2015uav,RN133,RN134,RN135,RN136,RN137}	]
				[\underline{\textit{Intelligent cloud computing services}} \\ Longest transmission delay\\ Strongest intelligence \\ \cite{luo2015uav,RN133,RN134,RN135,RN136,RN137}
				]]	
		\end{forest}}
		\caption{UAV-enabled intelligence to IoE}
		\label{fig:intelligent}
	\end{figure*}
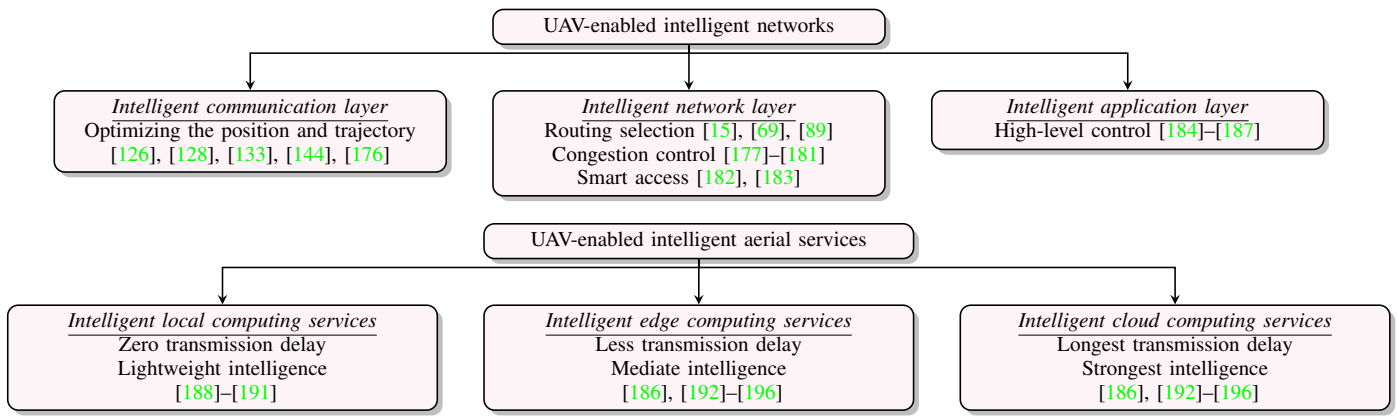

	\textit{(i) UAV-enabled intelligent networks:} UAVs can enable intelligent network functions to improve overall network performance. Similar to the network stack, we introduce three UAV-enabled intelligent network layers from bottom to up: the intelligent communication layer, the intelligent network layer, and the intelligent application layer.	
	
	The \emph{intelligent communication layer} relies on stable wireless connections between UAVs and IoE nodes while the stable and reliable communications require a good channel state, which can be achieved by deploying UAVs at optimal aerial positions. Section \ref{subsec: uavcommunication} has summarized the related optimization schemes, which discuss the issues such as the optimal trajectory and the topology for self-organized networks or relay-based networks.
	
	The \emph{intelligent network layer} can intelligently select optimized routing paths for data packets to improve the transmission efficiency and avoid congestion. Efficient routing relies on recognizing the source address and the destination address of every packet. In multi-UAV Ad Hoc networks, both the source address and the destination address are dynamically changed with flying multi-UAVs. Thus, the smart routing selection is adjusted with dynamic network topology (e.g., \cite{RN18}). Meanwhile, in UAV-aided communication networks, since data transmissions from multiple nodes to one node (i.e., from massive ground nodes to the UAV and from multiple UAVs to the ground access point) are not stable (may suffer from outage), the routing selection problem essentially becomes the problem of selecting right accessing UAVs to ensure the efficiency. 
	
	The \emph{intelligent application layer} can support diverse IoE applications via offering application programming interfaces (APIs). In this way, specific commands can be sent through APIs to control UAVs as well underlining layers (i.e. intelligent communication layer and intelligent network layer). The high-level commands can further guide UAVs to perform data collection operations, such as data compression and data aggregation.
	
	\textbf{Discussion.} Generally, due to the limited hardware resources, UAVs cannot support the network functions that require large storage space and intensive computational capabilities. Fortunately, the emerging network functional virtualization (NFV) and mobile edge computing (MEC) paradigms provide us with a solution to enable the above three intelligent network functions at UAVs. The solution is NFV/MEC-driven UAV-enabled intelligent network, in which NFV can virtualize UAV's network functions to programs and MEC can disperse the virtual function into the specific hardware at every mobile UAV. This solution can enable programmatically-efficient network configurations consisting of radio access, routing, switching, and firewalls. Practically, this NFV/MEC-driven solution is similar to a mobile software-defined network (SDN). By combining UAVs with SDN, the UAV network functions can be managed by SDN technologies, e.g., \cite{barritt2017operating}. 	
	
	\textit{(ii) UAV-enabled intelligent aerial services:} UAVs can provide the intelligent services by executing intelligent algorithms at UAVs as well as other distributed computing facilities as mentioned in Section~\ref{sec:ioe}. These intelligent algorithms guide UAVs to make smart decisions. Different UAV tasks (e.g., real-time monitoring, object tracking, and remote sensing) result in different decisions and different intelligent requirements. To enable various intelligent services, UAVs connecting with IoE can make full usage of local/edge/cloud computing facilities so as to enable local intelligent services, edge intelligent services, and cloud intelligent services. 
	
	To enable local intelligent services, UAVs require lightweight AI algorithms that can be running at UAVs with their own computing resources. Obviously, the local intelligent services can support immediate response with low delay while they are also restricted by the limited computing resources. In contrast, the edge intelligent services are provided at ground control stations or base stations. The stations generally have stronger computing capabilities as well as large storage space than UAVs while it may result in the inevitable delay between stations and UAVs.
	Furthermore, cloud computing servers with much stronger computing capabilities than edge servers so as to support the strongest intelligence while also bringing the longest transmission delay. In summary, UAVs can adopt local intelligence to make controls or immediate controls such as auto-control of flights and avoid obstacles. On the other hand, UAVs can harness the edge/cloud intelligence for time-tolerance decisions such as a remote photography and the long-distance good-delivery. 
	
	The corresponding studies on UAV-enabled intelligence of IoE are summarized as a tree graph as shown in Fig. \ref{fig:intelligent}.
	
	\subsubsection{UAV-enabled diversity to IoE} 
	\label{subsubsec:diverseueioe}
	We can utilize UAV technologies (as discussed in Section \ref{sec:uav}) to enable diverse applications of IoE. For instance, flexible UAV communication networks can be used to cover different geographic regions as mentioned in Section \ref{subsubsec:sclableueioe} so as to enable diverse geographical and stereoscopic applications. Additionally, combining with multiple intelligence methods (as mentioned in \ref{subsubsec:intellueioe}), UAVs are able to execute diverse intelligent algorithms in different scenarios (e.g., intelligent transportation systems, automatic package delivery, and aerial surveillance intelligence). Furthermore, with the prosperous development of IoE, different technologies can collaborate together to provide comprehensive services. We next introduce four typical Ue-IoE applications to show the integration of different technologies. 
	
	\textit{(i) Intelligent transportation system (ITS):} ITS as one of the major components of smart city is expected to automate transportation decisions through inter-connected vehicles as well as other transportation facilities (such as road side units and traffic control center). ITS aims to achieve more efficient decisions with low delays via automating ITS components including field support team, traffic police, road surveys, and rescue teams. In Ue-IoE, we can employ UAVs as the transportation information collector, the information transmitter, and even the executer for traffic schedule \cite{menouar2017uav}. Consequently, a set of UAVs can act as a field support team with high-efficiency, because they can promptly fly to the incident fields to conduct real-time reports and give emergency commands. Moreover, UAVs can be aerial traffic officers, when they fly over vehicles on a highway and detecting possible traffic violations. In ITS, UAV communication networks can provide temporary connections and UAVs can collect the real-time transportation information. Meanwhile, AI algorithms running at UAVs can make an immediate decision, consequently supporting intelligent information services such as conveying traffic information to road users.
	
	\textit{(ii) Rescue and Logistics:} We can use Ue-IoE to achieve on-line controls to UAVs for product delivery missions. In rural areas, the shortage of living materials is extremely urgent especially for the case of plagues or disasters. It is difficult to deliver urgent materials to these areas due to the disrupted roads. UAVs may serve as important carriers to deliver important materials such as medicines, foods, and clothes \cite{RN108,myslinski2019drone,evans2019drone,pizetta2019avoiding,shirani2019cooperative}. Meanwhile, dispatching UAVs for ordinary express delivery in urban areas has attracted lots of attention, thanks for flexible deployment and low cost of UAVs. Detailed studies for delivering UAVs can be found in previous literature. Referring to \cite{8600892}, a long-range and energy-efficient communication system for UAV delivery applications is developed, where LoRaWAN is used for semi-real-time telemetry. Moreover, \cite{7513397} proposes two multi-trip vehicle routing schemes for the cost-efficient drone delivery, in which the effect of battery and payload weight is considered for cost optimization. In addition, other studies for UAV delivery encompass hardware structure design \cite{7273724} and supplier cooperation \cite{8690828}.
	
	\textit{(iii) Aerial surveillance intelligence:} Over the last decades, there are a growing number of studies on using UAVs for aerial surveillance system (ASS) \cite{semsch2009autonomous, motlagh2017uav, kim2018designing}, in which UAVs mounted with the surveillance camera can employ controllable aerial photograph in a specified area. Different from fixed surveillance cameras that only offer fixed views on a limited region \cite{memos2018efficient}, UAV-enabled aerial surveillance can provide flexible views. Some studies investigate the autonomous capabilities of UAV-enabled surveillance, including regions detection \cite{yuan2017fire,nigam2009control}, objects positioning \cite{RN97}, and path planning \cite{nygaards2004navigation,faigl2019unsupervised,kim2018collision}. In addition, there is a strong demand to design an interface for connecting UAVs with the internet; this connection enables the Internet to monitor the UAV states. Hence, the intelligent algorithms can be adopted to analyze the surveillance photographs and videos. Some studies investigate the dedicated AI algorithms for rangeland inventory monitoring \cite{RN98}, endangered trees detecting in tropical rainforests \cite{RN102}, soil erosion monitoring \cite{RN83}, forest phenology monitoring \cite{RN87}.
	
	\textit{(iv) Unmanned military missions:}
	At the outset, UAVs are frequently used to perform military missions such as battlefield surveillance and attacks \cite{Handford2018Prospective}. UAVs can be employed as the aerial detectives in chaos battlefield, to detect the movements of enemy's troops and monitor the global battle situation \cite{6761569}. Moreover, UAVs can periodically spy the suspicious regions such as border surveillance \cite{abushahma2019comparative}. The UAV-aided relay communications in military have been widely investigated in existing literature (e.g., \cite{Orfanus2016Self, kim2017drone}). 	
	Obviously, UAV-enabled military missions (we name it unmanned military missions) save a lot of operating costs in manpower and fixed infrastructures. The unmanned missions must be conducted with reliable control to avoid exposure and crash. Especially for some missions in inaccessible areas (mountains, ice roads, deserts, etc.), the reliability of remote control is significant. Due to scalable coverage and intelligent computing resources, Ue-IoE is qualified to support flexible wireless connection and enable intelligent anti-detection ability. 

	The integration of UAVs with IoE can support diverse IoE applications own to the wide network coverage, big sharing database, and ubiquitous intelligence. Therefore, we will achieve high efficiency in every aspect of our daily life. 
	
	\section{Open research issues}
	\label{sec:issues}
	There are many technical issues that require dedicated investigations to fully realize UAV-enabled IoE. We outline five open issues in this promising area: i) the reasonable allocation methods to utilize the restricted resources in IoE so as to maximize the performance; ii) no security schemes to limit the illegal actions of UAVs; iii) absence of light-weight AI algorithms to autonomous mobility of UAVs; iv) no general framework of Ue-IoE to support heterogeneous compatible applications; v) absence of coordination schemes between various computing facilities (including UAVs, cloud and edge servers). 
	
	\subsection{Resource allocation}
	\label{subsec:allocation}
	Rational resource allocation (allocated resources contain energy supply, data storage, and computation capacity) for every node (including terminal nodes, UAVs, ground stations) can enhance the serving efficiency and reduce the cost \cite{yousafzai2019process,tang2019reconfigurable}. The future directions of the resource allocation can be divided into two categories: the global resource allocation (i.e., the number of nodes as well as the distribution or deployment of nodes in IoE) and the local resource allocation (i.e., the dedicated hardware configurations of every node in IoE). On the one hand, the global resource allocation focusing on the global high-efficiency of costs in time, energy, and the equipment. In UeIoE, the global efficiency can be optimized by deploying the various equipments in IoE, such as edge devices, cloud server \cite{RN8}, and the UAV \cite{wu2019fundamental}. While for the digital media transmission scenario in IoE, the global efficiency can be improved by networking algorithms \cite{stergiou2018algorithms} and video coding \cite{psannis2006impact}. On the other hand, the local resource allocation emphasizes on the dedicated efficiency in every task such as communication, computing, and data storage so as to decide the specified resource allocation in terms of power, data rate, duration of communication, data storage, computational complexity, computing time for communication tasks and computing tasks. For instance, in a local-network scenario (such as in a smart building), an efficient data collection algorithm is rather viable for improving the local data rate \cite{plageras2018efficient}. Additionally, for flying terminals - UAVs/drones, the energy management system is imperative to improve the using efficiency of limited energy \cite{RN30}. In a nutshell, Irrespective of various resource allocation strategies, the same principle always applies, i.e., the on-demand principle to fulfill expected requirements and coordinate resource consumption at different nodes. 
	
	\subsection{Security mechanism}
	\label{subsec:security}
	In anticipation of the growing IoE scale, the ubiquitous IoE communication will in face of security threats such as communication eavesdropping (or wiretapping) and node forgery. Take UAV-aided IoE communications as a typical example, they may suffer eavesdropping due to the openness of wireless medium. In addition, the illegal UAVs probably forge the identifications so as to connect with IoE. In spite of some recent studies analyzing UAVs' security \cite{javaid2012cyber,RN31}, there is still a long road to completely solve the security threats in Ue-IoE. With the increasing number of Ue-IoE applications, it is inevitable to propose and design security mechanisms that can address the above-mentioned threats. A research direction is to enhance the security level of UAV-based communications (including A2A links and A2G/G2A links) \cite{hua2019energy,kumar2017sdn}. Particularly, security countermeasures at physical layer or MAC layer need to be taken to mitigate eavesdropping and other malicious activities \cite{stergiou2018secure,bai2019energy}. Moreover, the improved authentication mechanisms can be adopted to avoid the illegal connections with IoE or UAVs \cite{punithavathi2019lightweight}. Furthermore, anti-fake or anti-forged UAVs mechanisms need to be leveraged to mitigate the node forgery issue.
	\subsection{Light-weight AI algorithms}
	\label{subsec:algorithm}
	To enhance the global intelligence of IoE, the light-weight AI algorithms are required to perform the immediate and precise responses at the devices with severe computation constraints (such as UAVs and sensors). Most of existing studies consider that AI algorithms used in the intelligent IoE applications have been executed at the cloud servers which have abundant computing resources. Therefore, the resource constraints have been ignored in those studies. For example, some studies present the AI algorithms to detect and count cars by using object detection algorithms enabled by deep convolutional neural networks (CNNs) \cite{ammour2017deep} or Faster R-CNN \cite{xu2017car}. Meanwhile, various UAV-based IoE applications such as human detection (i.e., finding pedestrians) \cite{liu2018extended}, weed mapping \cite{huang2018fully} have appeared. Even though these AI algorithms adopt the imagery records obtained by UAVs to conduct the mobile object recognition (detection), they have strict computing and storage requirements at computing facilities as most of them need to be executed or trained at superior computing servers (like GPU clusters). Therefore, it is necessary to design portable or tiny AI models dedicated for UAVs. For example, there are few studies on investigating low-complex algorithms for UAVs' navigation \cite{moskalenko2019model}, path-finding \cite{peng2019hybrid}, and moving-targets' tracking \cite{RN107}. The future IoE applications require lightweight AI algorithms to support autonomous intelligence with quick and precise responses.
	
	\subsection{Universal standard design}
	\label{subsec:framework}
	It is necessary to design a universal standard for IoE to orchestrate multiple heterogeneous ICT technologies as well as diverse IoE applications. The universalized Ue-IoE cannot only reduce the usage cost but also increase the serving efficiency. Referring to the network layer model in open system interconnection (OSI), the IoE universal standard should also include the specified rules for every IoE layer (i.e., a physical layer, network layer, and application layer). 
	For these three layers, the practical generality designs of IoE should include three aspects: the universal communication chips for end nodes, the universal network control protocols, and the universal computing facilities. However, existing studies only consider the hardware (chip sets) for communications at end nodes are only designed for particular communication protocols such as NB-IoT, MTC, LoRa-WAN \cite{8539671}. Moreover, the universal network protocols can be achieved by integrating SDN, but no global scheme was presented. The universal computing facilities can be achieved by integrating local, edge, and cloud computing facilities together. This integration is feasible by designing a high-level interface to connect the computing services between three types of computing facilities.
	
	\subsection{Coordination between various computing facilities}
	In the future, the coordination between edge and cloud computing facilities is an inevitable trend in order to cater for computation-intensive IoE applications \cite{stergiou2018secure}. These IoE applications require sufficient centralized AI decisions and distributive big data analytics to work seamlessly for real-time responses, such as smart manufacturing and intelligent transportation. The coordination is usually enabled by scheduling the computation tasks. Regarding computation-task scheduling, the edge computing facilities (such as UAVs and APs) first analyze the received computing requirements and then determine whether the task is required to upload to the remote cloud \cite{7901477}. 
	In practice, according to the practical requirements, some data preprocessed at the edge computing facilities need to be uploaded to the remote cloud servers for further storage and analytics, whereas other data can be computed at the edge computing facilities or at local nodes. . In this case, the computing requirements and the coordination mechanism may be specifically optimized to achieve global efficiency. In a long-term consideration, this optimization process is continuously adjustable in case of application upgrading and maintenance. Thus, the coordination of various computing resources (ranging from IoE nodes, UAVs, edge servers to remote cloud servers) is definitely a future direction.

	\section{Conclusion}
	\label{sec:conclusion}
	This paper aims at applying UAVs to enhance IoE's capabilities such as extended coverage, flexible intelligence, and more diverse applications. The envision of IoE in terms of three expectations (i.e., scalability, intelligence, and diversity) poses some realization challenges from coverage constraint, battery constraint, computing constraint, and security issues. With high mobility and flexible deployment, UAVs can help IoE to overcome these challenges. Therefore, we presented a comprehensive discussion on opportunities and solutions of UAVs in IoE.
	
	In particular, we first analysed the three expectations of IoE and presented an extensive survey of their enabling technologies. Moreover, we discussed the critical constraints/issues that hindered IoE's realization such as coverage constraint, battery constraint, computing constraint, and security issues. Furthermore, we presented an extensive review of UAV-related studies. The review mainly contains two aspects: the UAS design and the study on UAV communication networks. Accordingly, we have explored many opportunities of UAVs in IoE, i.e., the extension to ubiquitous connections, on-demand aerial intelligence, self-maintenance, power supply and sensor recycling, etc. Moreover, we presented a UAV-enabled IoE (namely Ue-IoE) solution by integrating UAVs with current ICT technologies. We demonstrated that Ue-IoE can greatly enhance the three expectations of IoE. We also presented three sub-solutions to fulfill the three expectations with summaries of corresponding studies. Finally, we outlined open issues as well as future directions in Ue-IoE. In summary, we have presented a comprehensive survey on the opportunities and challenges of using UAVs in IoE. This survey may serve as a research guideline for future studies on Ue-IoE.
	
	\balance
	\bibliographystyle{IEEEtran}
	\bibliography{ref}	
\end{document}